\documentclass[11pt]{myArticle}       

\usepackage{amsmath,amssymb,amsfonts} 
\usepackage{graphicx}
\usepackage{epsfig}
\usepackage{color}                    
\usepackage{myFancyhdr}               
\usepackage{makeidx}                  
\usepackage{helvet}                   
\usepackage{setspace}                 
\usepackage{fancybox}                 
\usepackage{float}                    
\usepackage{xspace}                   
\usepackage{wasysym}                  
\usepackage[numbers,sort&compress]{natbib}
\usepackage{hypernat}
\usepackage{wrapfig}
\usepackage{multirow}

\usepackage{hyperref}                 

\definecolor{mygray}{rgb}{0.3,0.32,0.35}
\definecolor{darkblue1}{rgb}{0,0,.2}
\definecolor{darkblue}{rgb}{0,0,.3}
\definecolor{darkred}{rgb}{0.5,0,0}
\pagecolor{white} 
\color{black}     
\hypersetup{breaklinks=true, 
            colorlinks=true, 
            linkcolor=darkblue1, 
            menucolor=darkblue1, 
            urlcolor=darkblue1,
            citecolor=darkblue1,
            pdftitle={The Heavy Object Tagger with Variable R},
            pdfauthor={Tobias Lapsien, Roman Kogler, Johannes Haller},
            pdfsubject={tagging with jet substructure},
            pdfkeywords={},
            pdfproducer={Tobias Lapsien, Roman Kogler, Johannes Haller}
}

\renewcommand{\headrulewidth}{0.5pt}    
\renewcommand{\footrulewidth}{0.5pt}    

\newcommand\allFontSize{\small}

\newenvironment{myquote}
               {\list{}{\leftmargin0cm}%
                \item\relax}
               {\endlist}

\usepackage[bf]{caption}

\newcommand\detailsSize{\allFontSize}
\newenvironment{details}%
{\begin{myquote}\vspace{-0.2cm}\detailsSize}{\end{myquote}\vspace{-0.2cm}}

\newfloat{codeexample}{H}{loc}
\floatname{codeexample}{\sf\allFontSize Code Example}

\newlength{\gfitterboxwidth}
\definecolor{DarkGray}{rgb}{0.4,0.42,0.45}
\definecolor{LightGray}{rgb}{0.97,0.98,0.98}

{\VerbatimEnvironment\normalsize
\setlength{\gfitterboxwidth}{\textwidth}\addtolength{\gfitterboxwidth}{-2.3\fboxsep}%
\begin{Sbox}\begin{minipage}[t]{\gfitterboxwidth}\begin{Verbatim}}%
{\end{Verbatim}\end{minipage}\end{Sbox}\vspace{1ex} \fcolorbox{DarkGray}{LightGray}{\TheSbox}}

\newfloat{option}{H}{loc}
\floatname{option}{\sf\allFontSize Option Table}

\usepackage{tabularx}
{\setlength{\gfitterboxwidth}{\textwidth}\addtolength{\gfitterboxwidth}{-2.3\fboxsep}%
\begin{Sbox}\begin{tabular*}{\gfitterboxwidth}{>{\rule[-0.5ex]{.0ex}{3.0ex}\tt\small}p{3cm}>{\tt\small}p{4.5cm}>{\small}p{5.7cm}}
&&\\[-0.7cm]
\rm\small Option  & \rm\small Values    & \rm\small Description\\[0.1cm]\hline
&&\\[-0.45cm]}%
{\\[-0.1cm]\end{tabular*}\end{Sbox}\vspace{1ex} \fbox{\TheSbox}}

{\setlength{\gfitterboxwidth}{\textwidth}\addtolength{\gfitterboxwidth}{-2.3\fboxsep}%
\begin{Sbox}\begin{tabular*}{\gfitterboxwidth}{>{\rule[-0.5ex]{.0ex}{3.0ex}\tt\small}p{3cm}>{\tt\small}p{3.5cm}>{\small}p{6.7cm}}
&&\\[-0.7cm]
\rm\small Option  & \rm\small Values    & \rm\small Description\\[0.1cm]\hline
&&\\[-0.45cm]}%
{\\[-0.1cm]\end{tabular*}\end{Sbox}\vspace{1ex} \fbox{\TheSbox}}

{\setlength{\gfitterboxwidth}{\textwidth}\addtolength{\gfitterboxwidth}{-2.3\fboxsep}%
\begin{Sbox}\begin{tabular*}{\gfitterboxwidth}{>{\rule[-0.5ex]{.0ex}{3.5ex}\tt\small}p{3.0cm}>{\tt\small}p{3.0cm}>{\small}p{7.2cm}}
&&\\[-0.7cm]
\rm\small Option  & \rm\small Values    & \rm\small Description\\[0.1cm]\hline
&&\\[-0.45cm]}%
{\\[-0.1cm]\end{tabular*}\end{Sbox}\vspace{1ex} \fbox{\TheSbox}}

{\setlength{\gfitterboxwidth}{\textwidth}\addtolength{\gfitterboxwidth}{-2.3\fboxsep}%
\begin{Sbox}\begin{tabular*}{\gfitterboxwidth}{>{\rule[-0.5ex]{.0ex}{3.0ex}\tt\small}p{2.5cm}>{\tt\small}p{3.5cm}>{\small}p{7.2cm}}
&&\\[-0.7cm]
\rm\small Option  & \rm\small Values    & \rm\small Description\\[0.1cm]\hline
&&\\[-0.45cm]}%
{\\[-0.1cm]\end{tabular*}\end{Sbox}\vspace{1ex} \fbox{\TheSbox}}

{\setlength{\gfitterboxwidth}{\textwidth}\addtolength{\gfitterboxwidth}{-2\fboxsep}%
\begin{Sbox}\begin{tabular*}{\gfitterboxwidth}{>{\rule[-0.5ex]{.0ex}{3.0ex}\tt\small}p{4.5cm}>{\tt\small}p{3.2cm}>{\small}p{5.5cm}}
&&\\[-0.7cm]
\rm\small Option  & \rm\small Values    & \rm\small Description\\[0.1cm]\hline
&&\\[-0.45cm]}%
{\\[-0.1cm]\end{tabular*}\end{Sbox}\vspace{1ex} \fbox{\TheSbox}}

{\setlength{\gfitterboxwidth}{\textwidth}\addtolength{\gfitterboxwidth}{-2\fboxsep}%
\begin{Sbox}\begin{tabular*}{\gfitterboxwidth}{>{\rule[-0.5ex]{.0ex}{3.0ex}\tt\small}p{4.0cm}>{\tt\small}p{3.0cm}>{\small}p{6.2cm}}
&&\\[-0.7cm]
\rm\small Option  & \rm\small Values    & \rm\small Description\\[0.1cm]\hline
&&\\[-0.45cm]}%
{\\[-0.1cm]\end{tabular*}\end{Sbox}\vspace{1ex} \fbox{\TheSbox}}

\newfloat{programs}{H}{loc}
\floatname{programs}{\sf Table}

\usepackage{tabularx}
{\setlength{\gfitterboxwidth}{\textwidth}
\addtolength{\gfitterboxwidth}{-2\fboxsep}%
\begin{Sbox}\begin{tabular*}{\gfitterboxwidth}
{>{\rule[-0.5ex]{.0ex}{3.0ex}\tt\small}p{4.1cm}>{\small}p{9.5cm}}&\\[-0.7cm]
\rm\small Macro & \rm\small Description\\[0.1cm]\hline&\\[-0.45cm]}%
{\\[-0.1cm]\end{tabular*}\end{Sbox}\vspace{1ex} \fbox{\TheSbox}}

\newenvironment{parametertable}%
{\setlength{\gfitterboxwidth}{0.75\textwidth}%
\begin{tabular*}{\gfitterboxwidth}{>{\rule[-0.5ex]{.0ex}{3.0ex}\small}p{2cm}>{\small}p{2.5cm}>{\small}p{9.7cm}}
&&\\[-0.7cm]
\rm\small Parameter  & \rm\small Default  & \rm\small Description\\[0.1cm]\hline
&&\\[-0.45cm]}%
{\\[-0.1cm]\end{tabular*}\vspace{1ex}}

{\setlength{\gfitterboxwidth}{0.95\textwidth}%
\begin{tabular*}{\gfitterboxwidth}
{>{\rule[-0.5ex]{.0ex}{3.0ex}\small}p{1.8cm}>{\small}p{8cm}>{\small}p{3.5cm}}
&&\\[-0.7cm]
\rm\small Parameter  & \rm\small Description  & \rm\small Selection\\[0.1cm]\hline
&&\\[-0.45cm]}%
{\\[-0.1cm]\end{tabular*}\vspace{1ex}}

\newcommand{\pt}        {\ensuremath{p_\mathrm{T}}\xspace}
\newcommand{\fpt}       {\ensuremath{f_{p_\mathrm{T}}\xspace}}
\newcommand{\ptsub}     {\ensuremath{p_\mathrm{T,sub}}\xspace}
\newcommand{\ptone}     {\ensuremath{p_\mathrm{T,1}}\xspace}
\newcommand{\pti}     {\ensuremath{p_{\mathrm{T},i}}\xspace}
\newcommand{\ptj}     {\ensuremath{p_{\mathrm{T},j}}\xspace}
\newcommand{\ptij}     {\ensuremath{p_{\mathrm{T}i,j}}\xspace}

\newcommand{\kt}        {\ensuremath{k_t}\xspace}
\newcommand{\Fastjet}   {\textsc{FastJet}\xspace}

\newcommand{\HOTVR}   {\textsc{HOTVR}\xspace}
\newcommand{\HEPTT}   {\textsc{HEPTopTagger}\xspace}
\newcommand{\dij}     {\ensuremath{d_{ij}}\xspace}
\newcommand{\diB}     {\ensuremath{d_{i\mathrm{B}}}\xspace}
\newcommand{\Rmin}    {\ensuremath{R_\mathrm{min}}\xspace}
\newcommand{\Rmax}    {\ensuremath{R_\mathrm{max}}\xspace}
\newcommand{\Reff}    {\ensuremath{R_\mathrm{eff}}\xspace}
\newcommand{\mjet}    {\ensuremath{m_\mathrm{jet}}\xspace}
\newcommand{\mmin}    {\ensuremath{m_\mathrm{min}}\xspace}
\newcommand{\Nsub}    {\ensuremath{N_\mathrm{sub}}\xspace}
\newcommand{\ftau}    {\ensuremath{\tau_{3/2}}\xspace}
\newcommand{\es}      {\ensuremath{\varepsilon_{S}}\xspace}
\newcommand{\eb}      {\ensuremath{\varepsilon_{B}}\xspace}
\newcommand{\Rfiltmax}{\ensuremath{R_{\mathrm{filt}}^{\mathrm{max}}}\xspace}
\newcommand{\Nfilt}   {\ensuremath{N_{\mathrm{filt}}}\xspace}
\newcommand{\mrec}    {\ensuremath{m_{\mathrm{rec}}}\xspace}
\newcommand{\mrecf}   {\ensuremath{m_{\mathrm{rec}}^{1.5}}\xspace}

\newcommand{\tq}    {\ensuremath{\mathrm{t}}\xspace}

\newcommand{\ttbar} {\ensuremath{\mathrm{t}\overline{\mathrm{t}}}\xspace}

\newcommand{\pp}    {\ensuremath{\mathrm{p}\mathrm{p}}\xspace}
\newcommand{\W} {\ensuremath{\mathrm{W}}\xspace}

\newcommand{\Z} {\ensuremath{\mathrm{Z}}\xspace}

\newcommand{\Hig} {\ensuremath{\mathrm{H}}\xspace}
\mathchardef\Upsilon="7107
\def\Y#1S{\ensuremath{\Upsilon{(#1S)}}\xspace}

\newcommand{\Kbar    }{\kern 0.2em\overline{\kern -0.2em K}{}\xspace}

\newcommand{\Kz      }{\ensuremath{K^0}\xspace}
\newcommand{\Kzb     }{\ensuremath{\Kbar^0}\xspace}
\newcommand{\KzKzb   }{\ensuremath{\Kz \kern -0.16em \Kzb}\xspace}
\newcommand{\Kp      }{\ensuremath{K^+}\xspace}
\newcommand{\Km      }{\ensuremath{K^-}\xspace}

\newcommand{\KpKm    }{\ensuremath{\Kp \kern -0.16em \Km}\xspace}

\newcommand\Dbar    {\kern 0.18em\overline{\kern -0.18em D}{}\xspace}

\newcommand\Bbar    {\kern 0.18em\overline{\kern -0.18em B}{}\xspace}

\newcommand\Bz      {\ensuremath{B^0}\xspace}

\newcommand\Bzb     {\ensuremath{\Bbar^0}\xspace}
\newcommand\Bu      {\ensuremath{B^+}\xspace}
\newcommand\Bub     {\ensuremath{B^-}\xspace}

\newcommand\BpBm    {\ensuremath{\Bu {\kern -0.16em \Bub}}\xspace}
\newcommand\Bs      {\ensuremath{B^0_{s}}\xspace}
\newcommand\Bsb     {\ensuremath{\Bbar^0_{s}}\xspace}

\newcommand\BzBzb   {\ensuremath{\Bz {\kern -0.16em \Bzb}}\xspace}
\newcommand\BszBszb {\ensuremath{\Bs {\kern -0.16em \Bsb}}\xspace}

\newcommand{\Order}{\ensuremath{{\cal O}}\xspace}

\newcommand{\tev}{\ensuremath{\,\mathrm{Te\kern -0.1em V}}\xspace}
\newcommand{\gev}{\ensuremath{\,\mathrm{Ge\kern -0.1em V}}\xspace}
\newcommand{\mev}{\ensuremath{\,\mathrm{Me\kern -0.1em V}}\xspace}
\newcommand{\kev}{\ensuremath{\,\mathrm{ke\kern -0.1em V}}\xspace}
\newcommand{\ev}{\ensuremath{\,\mathrm{e\kern -0.1em V}}\xspace}
\newcommand{\gevc}{\ensuremath{\,{\mathrm{Ge\kern -0.1em V\!/}c}}\xspace}
\newcommand{\mevc}{\ensuremath{\,{\mathrm{Me\kern -0.1em V\!/}c}}\xspace}
\newcommand{\gevcc}{\ensuremath{\,{\mathrm{Ge\kern -0.1em V\!/}c^2}}\xspace}
\newcommand{\mevcc}{\ensuremath{\,{\mathrm{Me\kern -0.1em V\!/}c^2}}\xspace}

\newcommand{\bei}{\begin{itemize}}
\newcommand{\eei}{\end{itemize}}
\newcommand{\beq}{\begin{equation}}
\newcommand{\eeq}{\end{equation}}
\newcommand{\beqn}{\begin{eqnarray}}
\newcommand{\eeqn}{\end{eqnarray}}
\newcommand{\beqns}{\begin{eqnarray*}}
\newcommand{\eeqns}{\end{eqnarray*}}
\newcommand{\bitm}{\begin{itemize}}
\newcommand{\eitm}{\end{itemize}}

\def\@citex[#1]#2{\if@filesw\immediate\write\@auxout{\string\citation{#2}}\fi
  \@tempcnta\z@\@tempcntb\m@ne\def\@citea{}\@cite{\@for\@citeb:=#2\do
    {\@ifundefined
       {b@\@citeb}{\@citeo\@tempcntb\m@ne\@citea
        \def\@citea{,\penalty\@m\ }{\bf ?}\@warning
       {Citation `\@citeb' on page \thepage \space undefined}}%
    {\setbox\z@\hbox{\global\@tempcntc0\csname b@\@citeb\endcsname\relax}%
     \ifnum\@tempcntc=\z@ \@citeo\@tempcntb\m@ne
       \@citea\def\@citea{,\penalty\@m}
       \hbox{\csname b@\@citeb\endcsname}%
     \else
      \advance\@tempcntb\@ne
      \ifnum\@tempcntb=\@tempcntc
      \else\advance\@tempcntb\m@ne\@citeo
      \@tempcnta\@tempcntc\@tempcntb\@tempcntc\fi\fi}}\@citeo}{#1}}

\def\@citeo{\ifnum\@tempcnta>\@tempcntb\else\@citea
  \def\@citea{,\penalty\@m}%
  \ifnum\@tempcnta=\@tempcntb\the\@tempcnta\else
   {\advance\@tempcnta\@ne\ifnum\@tempcnta=\@tempcntb \else
\def\@citea{--}\fi
    \advance\@tempcnta\m@ne\the\@tempcnta\@citea\the\@tempcntb}\fi\fi}

\xspace

\makeindex
\begin{document}


%
%
\pagenumbering{arabic}
{\small
\color{mygray}
\begin{flushright}
{\sf\em arXiv:1606.04961} \\
{\sf\em October 19, 2016} \\
\def\UrlFont{\sf\em}
\end{flushright}
}
\def\UrlFont{\rm}

\vspace{1.3cm}

{\sf\LARGE\bfseries
\begin{center}
A new tagger for hadronically decaying \\[0.2cm]
heavy particles at the LHC
\end{center}
}

\vspace{2.0cm}

{\large
T.~Lapsien, R.~Kogler, J.~Haller
}

\vspace{0.2cm}

\begin{details}
  Institut f\"ur Experimentalphysik, Universit\"at Hamburg, Germany\\
\end{details}

\vspace{2.0cm}

\begin{details} {\sf\bfseries Abstract} 
  --- A new algorithm for the identification of boosted,
  hadronically decaying, heavy particles at the LHC is presented.  The
  algorithm is based on the known procedure of jet clustering with variable distance
  parameter $R$ and adapts the jet size to its transverse momentum \pt.
  Subjets are found using a mass jump condition.  The resulting
  algorithm -- called Heavy Object Tagger with Variable $R$ (HOTVR) --
  features little algorithmic complexity and combines jet clustering, subjet finding and rejection of soft
  clusters in one sequence.  
  While the HOTVR algorithm can be used for the identification of  
  any heavy object decaying hadronically, e.g.\ W, Z, H, t, or possible
  new heavy resonances, this paper targets specifically the tagging of boosted top
  quarks.  
  The studies presented here demonstrate a stable performance of
  the HOTVR algorithm in a wide range of top quark \pt, from low \pt, where the
  decay products can be resolved, to the region of boosted
  decays at high \pt.
\end{details}

\thispagestyle{empty}

\newpage
%
%
\section{Introduction}
\label{sec:intro}

The identification of hadronically decaying heavy Standard
Model (SM) particles (\W, \Z, \Hig, \tq) is an important ingredient in an increasing number of 
SM analyses and searches for new physics at the LHC. For a particle with high
energy, the large Lorentz factor leads to decay products which are
collimated in the laboratory rest frame and result in a single jet.
The task of separating these decays from the vast amount of
background from QCD multijet production has been approached with a variety of
jet substructure developments in recent years~\cite{Butterworth:2008iy,
Ellis:2009su, Salam:2009jx,
  Abdesselam:2010pt, Altheimer:2012mn, Altheimer:2013yza,
  Adams:2015hiv}.
The techniques face the challenge of a stable performance in significantly different 
kinematic regimes: from the region of low transverse momentum \pt, where the decay products 
can be resolved, to the boosted regime of high \pt.

The existing algorithms can be classified into two approaches. 
The bottom-up approach extrapolates from the resolved into the 
boosted regime by successively combining small radius jets, similar to 
an exclusive jet clustering (e.g. the JADE algorithm~\cite{jade1, jade2}).
Modern algorithms have been devised for the task of 
heavy object identification; examples are the collection of jets in
buckets~\cite{Buckley:2013lpa, Buckley:2013auc} or the recently
proposed XCone algorithm~\cite{Stewart:2015waa, Thaler:2015xaa}. These methods combine manageable
complexity with promising performance, but have not been studied in
experimental analyses so far.
The larger number of algorithms follows the top-down approach which 
starts from large radius jets followed by subsequent declustering steps. 
These algorithms are based on jet clustering with a fixed distance parameter $R$, where 
jet grooming methods like filtering~\cite{Butterworth:2008iy}, pruning~\cite{pruning},
trimming~\cite{trimming} or soft drop~\cite{softdrop} are used to remove
soft radiation and contributions from the underlying event such that substructure observables like
the jet mass reflect the hard underlying process.
Alternatively the variable $R$ jet algorithm~\cite{Krohn:2009zg} 
can be used to dynamically reduce the jet distance parameter with increasing \pt of the decaying particle. 
The algorithm was used in studies of new heavy resonances decaying 
to final states with two and four gluons~\cite{trimming}, 
and also top quark, \W and Higgs decays at LHC energies~\cite{ATL-PHYS-PUB-2016-013}, 
and a similar algorithm at energies of a future hadron collider~\cite{Larkoski:2015yqa}. 
Additional substructure information like the $\kt$ splitting
scales~\cite{Butterworth:2002tt}, 
N-subjettiness~\cite{Stewart:2010tn, Thaler:2010tr, Thaler:2011gf}, 
energy correlation functions~\cite{Larkoski:2013eya} or Qjets~\cite{qjets} 
are often used to further improve the performance of substructure algorithms.
Combinations of these methods are used for the tagging of top
quarks~\cite{johnhopkins, cmstt, htt2009, htt2010, htt2015}, 
where also more theoretically motivated taggers have been proposed~\cite{sd2011, sd2012}.
The ATLAS and CMS collaborations have commissioned a number of
the techniques mentioned above and studied their behaviour~\cite{ATLAS:2012am,
  Aad:2012meb, Aad:2013gja, Chatrchyan:2013vbb, JME-13-007,
  Khachatryan:2014vla, Aad:2015rpa, Aad:2016pux}.  
Several top-tagging algorithms have been employed
successfully in various searches for new
physics~\cite{Chatrchyan:2012ypy, Chatrchyan:2012rva,
  Chatrchyan:2012ku, Aad:2012ans, Chatrchyan:2013wfa,
  Chatrchyan:2013lca, Aad:2013nca, Khachatryan:2014gha,
  Aad:2014xra,Aad:2014xka, Khachatryan:2015wza, Khachatryan:2015sma,
  Khachatryan:2015oba, Khachatryan:2015gza, Khachatryan:2015axa,
  Khachatryan:2015mta, Khachatryan:2015ywa, Khachatryan:2015bma,
  Aad:2015owa,
  Aad:2015ufa,Aad:2015fna,Aad:2015uka,Aad:2015voa,Aad:2015dva,Aad:2015ipg,
  Khachatryan:2016cfa, Khachatryan:2016yji, Aad:2016qpo, Aad:2016shx}
and in SM top quark measurements~\cite{Aad:2015hna,Khachatryan:2016gxp} in
LHC analyses with Run~1 data.
 
At LHC's Run 2 the production rates of particles with high \pt 
have increased 
and the importance of boosted analyses is further enhanced. 
Modifications of existing taggers have been proposed and
studied in simulation~\cite{JME-14-002, JME-15-002, ATL-PHYS-PUB-2015-033,
  ATL-PHYS-PUB-2015-035}.  In most cases, a modest performance
improvement is contrasted with a significantly increased algorithmic complexity. 
A simple but robust algorithm~\cite{ATL-PHYS-PUB-2015-053} proposed by 
the ATLAS collaboration has a slightly reduced performance.
In addition, recent developments of top-down taggers aim at closing the gap between 
the resolved and the boosted regime with rather complex algorithmic procedures using several 
clustering, declustering, mass drop and filtering steps. 
Examples are the scale invariant 
tagger~\cite{Gouzevitch:2013qca} and the \HEPTT in OptimalR mode~\cite{htt2015}.

In this work we introduce a new tagger 
useful in the resolved, the transition and the boosted regime,
achieved with only little algorithmic complexity.
The tagger is based on the variable $R$ jet algorithm~\cite{Krohn:2009zg}, 
which adapts the jet distance parameter
dynamically to the \pt of the boosted object.  
A mass jump condition~\cite{Stoll:2014hsa, Hamaguchi:2015uqa} is included in
the clustering process, which forms subjets reflecting the dynamics of
the underlying hard decay, and enables efficient background
suppression. The resulting Heavy Object Tagger with Variable $R$
(\HOTVR) accommodates jet clustering, subjet finding and the rejection
of soft radiation in one sequence, without the need of 
declustering and following grooming steps.
In this paper we demonstrate the algorithm's properties and characteristics in 
hadronic top decays and leave studies of the decays 
of \W, \Z, \Hig and possible new resonances to future work.

The paper is organised as follows. In section~\ref{sec:algorithm} the
\HOTVR algorithm is described.  Its characteristics, free parameters
and their influence on the jet and subjet clustering, the collinear
and infrared safety and timing performance are discussed in
section~\ref{sec:props}.  The algorithm's performance for hadronic top quark
quark decays and a comparison with other commonly used taggers is
presented in section~\ref{sec:performance}. A conclusion is given in
section~\ref{sec:conclusion}.


\section{The algorithm}
\label{sec:algorithm}

The \HOTVR algorithm is based on the variable $R$ (VR) jet
algorithm~\cite{Krohn:2009zg}.  Like all sequential recombination
algorithms, it starts with an input list of pseudojets\footnote{ We use the
  notation pseudojet to denote entities entering the jet clustering.
  These can be partons, stable particles, reconstructed detector
  objects or combined objects from a previous clustering iteration.  }
and continues the processing until the input list is empty.  The
algorithm uses the distance measures \dij and \diB, defined as
\begin{align}
  \dij  &= \min\left[ p_{\mathrm{T},i}^{2n},p_{\mathrm{T},j}^{2n} \right] \Delta R_{ij}^2 \, ,\label{eq:dij}\\
  \diB  &= p_{\mathrm{T},i}^{2n} \, \Reff^2(p_{\mathrm{T},i}) \, , \label{eq:dib}\\
  \Reff(\pt) &= \frac{\rho}{\pt} \, . \label{eq:reff}
\end{align}
The value of \dij can be interpreted as distance between two
pseudojets $i$ and $j$, where $p_{\mathrm{T},i}$ is the transverse
momentum of pseudojet $i$ and $\Delta R_{ij} = \sqrt{(y_i - y_j)^2 +
  (\phi_i - \phi_j)^2}$ is the angular distance in rapidity $y$ and
azimuth $\phi$ between the pseudojets $i$ and $j$.  
The value of \diB denotes the distance between
pseudojet $i$ and the beam. For a fixed distance parameter of 
$\Reff = R$ in Eq.~\eqref{eq:reff}, 
the anti-\kt~\cite{Cacciari:2008gp},
Cambridge/Aachen (CA)~\cite{ca1,ca2} and \kt~\cite{kt1, kt2}
algorithms are obtained for the choices $n = -1, 0, 1$,
respectively. 
For the \HOTVR algorithm $n=0$ is used, corresponding to CA clustering.  
However, in the VR algorithm \Reff is an
effective distance parameter, which scales with $1/\pt$ (cf.\ 
Eq.~\eqref{eq:reff}) leading to broader jets at low \pt and narrower jets at high
\pt. 
The scale $\rho$ determines the slope of \Reff.
For robustness of the algorithm with respect to experimental effects 
a minimum and a maximum cut-off for \Reff is introduced,
\begin{equation}
\Reff =
\begin{cases}
\Rmin   & \text{for } \rho/\pt < \Rmin \, ,  \\
\Rmax & \text{for } \rho/\pt > \Rmax \, , \\
\rho/\pt  & \text{else} \, . 
\end{cases}
\end{equation}

A known shortcoming of the VR algorithm is the clustering of
additional radiation into jets in QCD multijet production, resulting
in a higher jet \pt on average and an increased rate 
once a \pt selection is applied~\cite{Krohn:2009zg}.
The \HOTVR algorithm approaches this issue by modifying the jet
clustering procedure with a veto based on the invariant mass of the
pseudojet pair, inspired by the recently proposed mass jump
algorithm~\cite{Stoll:2014hsa}. The mass jump veto prevents the
recombination of two pseudojets $i$ and $j$ if the combined invariant
mass $m_{ij}$ is not large enough,
\begin{equation}
\theta\cdot m_{ij}>\max\left[m_i,m_j\right] \, .
\label{eq:massjump}
\end{equation}
The parameter $\theta$ determines the strength of the mass jump veto
and can be chosen from the interval $[0,1]$. 
The mass jump criterion~\eqref{eq:massjump} is only 
applied if the mass $m_{ij}$ is larger than a mass threshold $\mu$
\begin{equation}
m_{ij} > \mu \, .
\end{equation} 
In case a mass jump is found and the \pt of the 
pseudojets $i$ and $j$ fulfil
\begin{equation}
\ptij > \ptsub 
\end{equation} 
the pseudojets are combined.
The resulting pseudojet enters the next
clustering step and the initial pseudojets are stored as separate 
subjets.
In case the mass jump criterion is not fulfilled or the pseudojets
are softer than \ptsub, the lighter pseudojet or the one too soft is removed from
the list. This step reduces the effect of additional
activity (soft radiation, underlying event, pile-up) and effectively
stabilises the jet mass over a large range of \pt.

The full \HOTVR algorithm can be summarised as follows.
\begin{itemize}
\item[1)] If the smallest distance parameter is \diB, store the
  pseudojet $i$ as jet and remove it from the input list of pseudojets.
\item[2)] If the smallest distance parameter is \dij and $m_{ij} \leq \mu$, combine 
  $i$ and $j$.
\item[3)] If the smallest distance parameter is \dij and $m_{ij} >
  \mu$, check the mass jump criterion $\theta\cdot m_{ij} >
  \max[m_i,m_j]$.
 \begin{itemize}
 \item[a)] If the mass jump criterion is not fulfilled, compare the
   masses of the two pseudojets and remove the one with the lower mass from the
   input list.
 \item[b)] If the mass jump criterion is fulfilled, 
     check the transverse momenta of the subjets $i$ and $j$. 
\begin{itemize}
\item[i)] If $\pti < \ptsub$ or $\ptj < \ptsub$, remove the respective pseudojet from the input list.
\item[ii)] Else, combine pseudojets $i$ and $j$.  
   Store the pseudojets $i$ and $j$ as subjets of the
   combined pseudojet.  In case $i$ or $j$ have already subjets,
   associate their subjets with the combined pseudojet.
\end{itemize}
 \end{itemize}
\item[4)] Continue with 1) until the input list of pseudojets is empty.
\end{itemize}
The algorithm results in jets with an effective size depending on $\pt$
and associated subjets. It incorporates jet finding, subjet finding
and the rejection of soft radiation in one clustering
sequence.  

The algorithm is available as plugin to \Fastjet~\cite{Cacciari:2005hq, Fastjet} 
and can be obtained through the \Fastjet Contribs package~\cite{fjcontribs}. 
Its implementation is based on the implementations of the mass jump and
VR algorithms in the \Fastjet Contribs packages \textsc{ClusteringVetoPlugin} 
1.0.0 and \textsc{VariableR} 1.1.1, respectively.
These implementations have been adapted and modified to make 
the \HOTVR software an independent \Fastjet plugin.

\section{Characteristics and properties}
\label{sec:props}

\subsubsection*{Parameters, jet and subjet finding} 

\begin{table}[tb]
\centering
\begin{parametertable}
\Rmin & 0.1 & Minimum value of \Reff. \\
\Rmax & 1.5  & Maximum value of \Reff.  \\
$\rho$ & 600\gev & Slope of \Reff.\\
$\mu$ & 30\gev & Mass jump threshold. \\
$\theta$ & 0.7 & Mass jump strength. \\
$\ptsub$ & 30\gev & Minimum \pt of subjets.
\end{parametertable}
\caption{\label{tab:VRhot_params} Parameters of the \HOTVR algorithm. The default values
are given for the top-tagging mode. }
\end{table}

In total, the algorithm has six parameters, which are
listed in Tab.~\ref{tab:VRhot_params}. While the first three
parameters steer the VR part of the algorithm, the last three
define the mass jump condition. The default values given in the table
have been optimised for top quark tagging in $\mathrm{pp}$ collisions at
$\sqrt{s}$=13\tev.

The original VR algorithm is recovered for $\mu\to\infty$. 
In this case, for $\rho \to 0$ the algorithm is identical to the 
CA algorithm with a distance parameter of $R = \Rmin$. 
Similarly, for $\rho \to \infty$ the CA algorithm is obtained with $R = \Rmax$. 
For values of $\rho$ corresponding to the typical scale of an event
($m$ or $\pt$ in the range of $\mathcal{O}(100\gev)$) jets
are clustered with an adaptive distance parameter between $\Rmin$ and $\Rmax$. 
Higher values of $\rho$ result in larger jet sizes.

The number of subjets found is modified by the mass jump parameters $\mu$, 
$\theta$ and \ptsub. 
Once the pseudojets become sufficiently heavy due to clustering, the mass jump 
threshold $\mu$ results in a rejection of soft and light pseudojets.
For a fixed value of $\mu$, the strength of this jet grooming depends on the parameters $\theta$ and \ptsub.
For $\theta=1$ the condition \eqref{eq:massjump} is always fulfilled
and no pseudojets are rejected (equivalent to the case $\mu\to\infty$). 
Conversely, the case of $\theta=0$ 
results in a VR jet clustering which stops as soon as a jet mass of $\mu$ is reached. 
The algorithm results in subjets with a maximum mass of $\mu$.
Additional jet grooming is obtained by setting $\ptsub>0$. This results in
subjets with a minimum \pt of \ptsub, effectively removing soft radiation 
and improving the tagging performance at small \pt of the heavy object.

\begin{figure}[tbp] 
  \centering
  \includegraphics[width=0.47\textwidth]{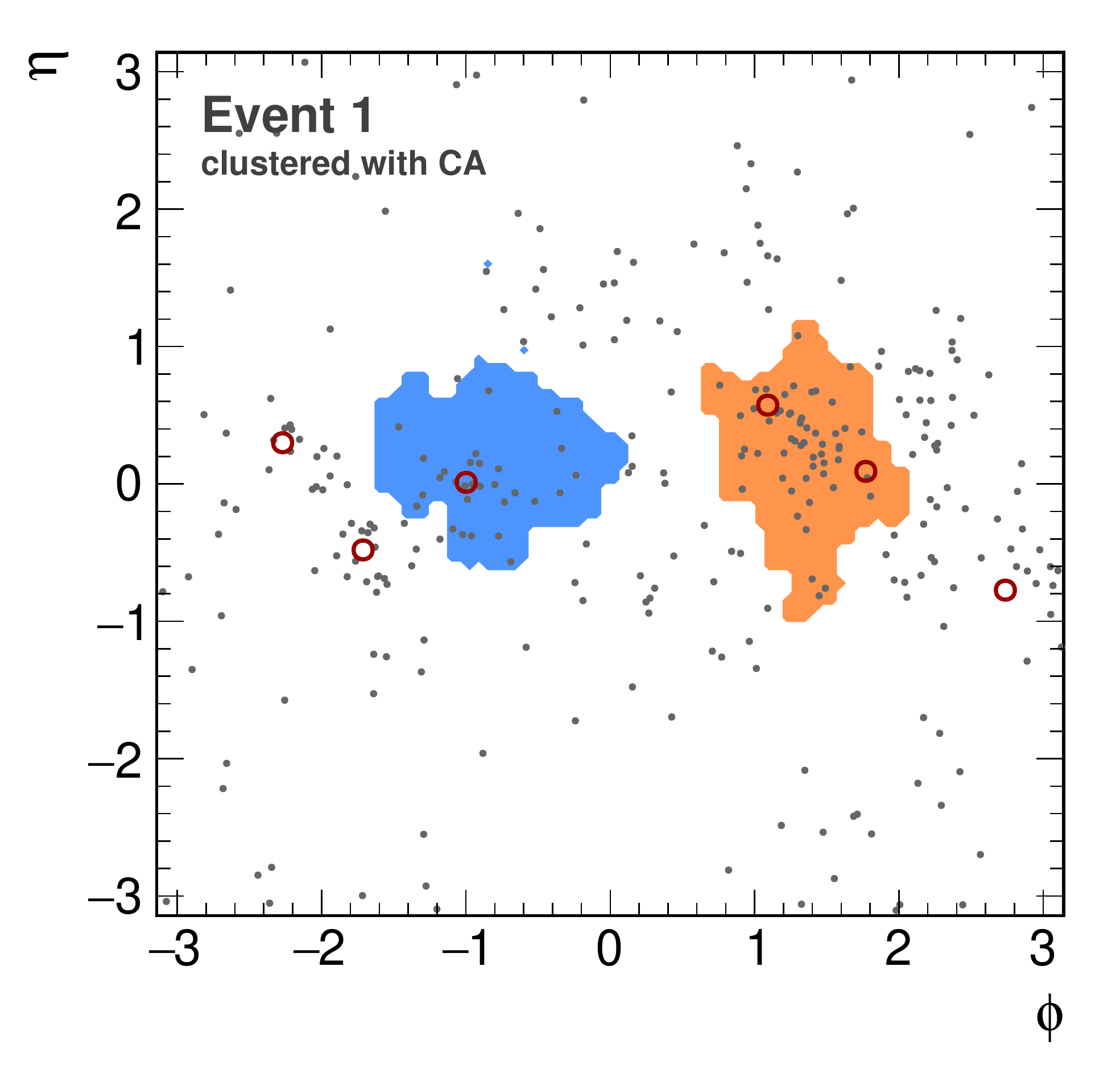}
  \includegraphics[width=0.47\textwidth]{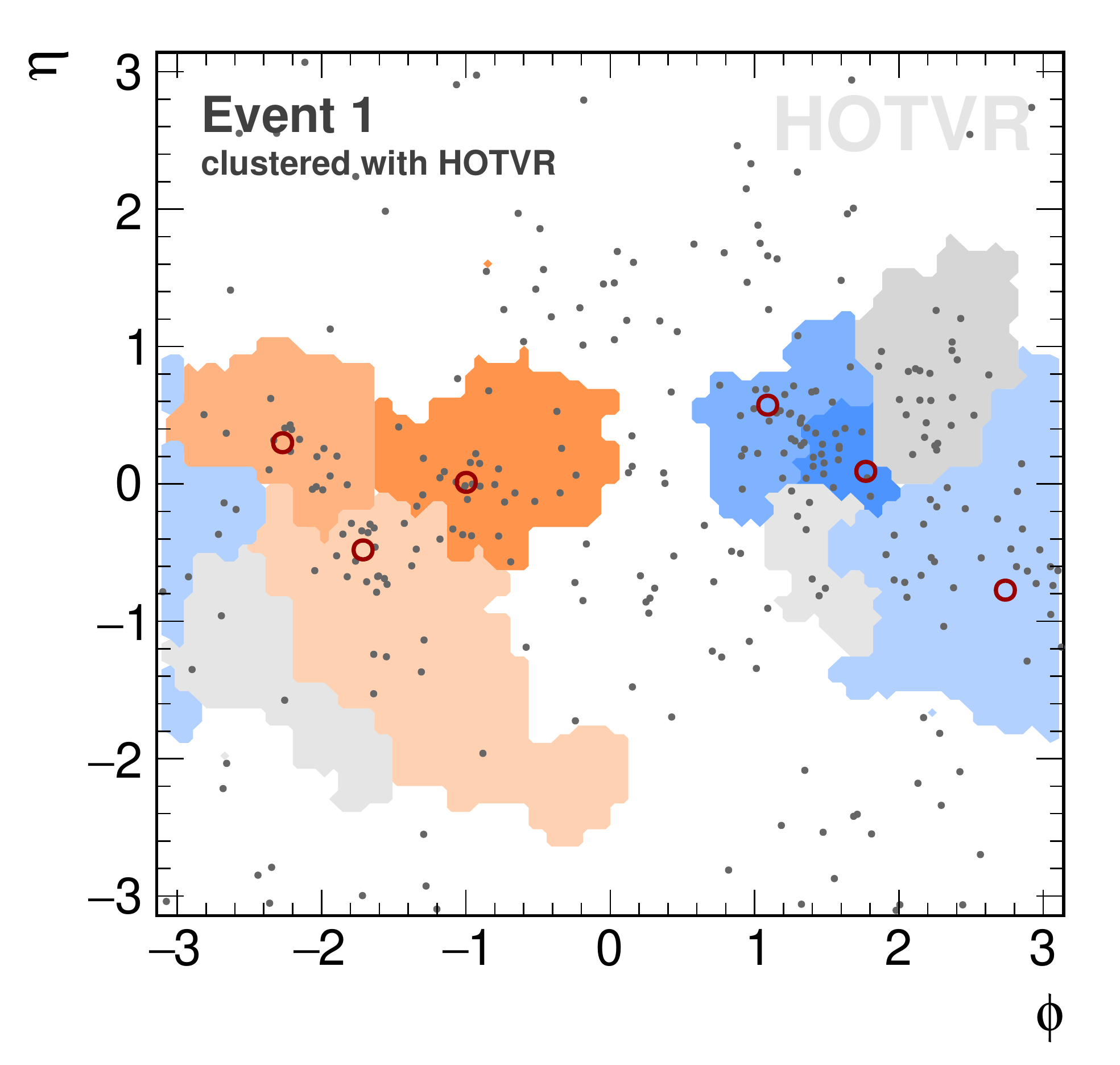}
  \includegraphics[width=0.47\textwidth]{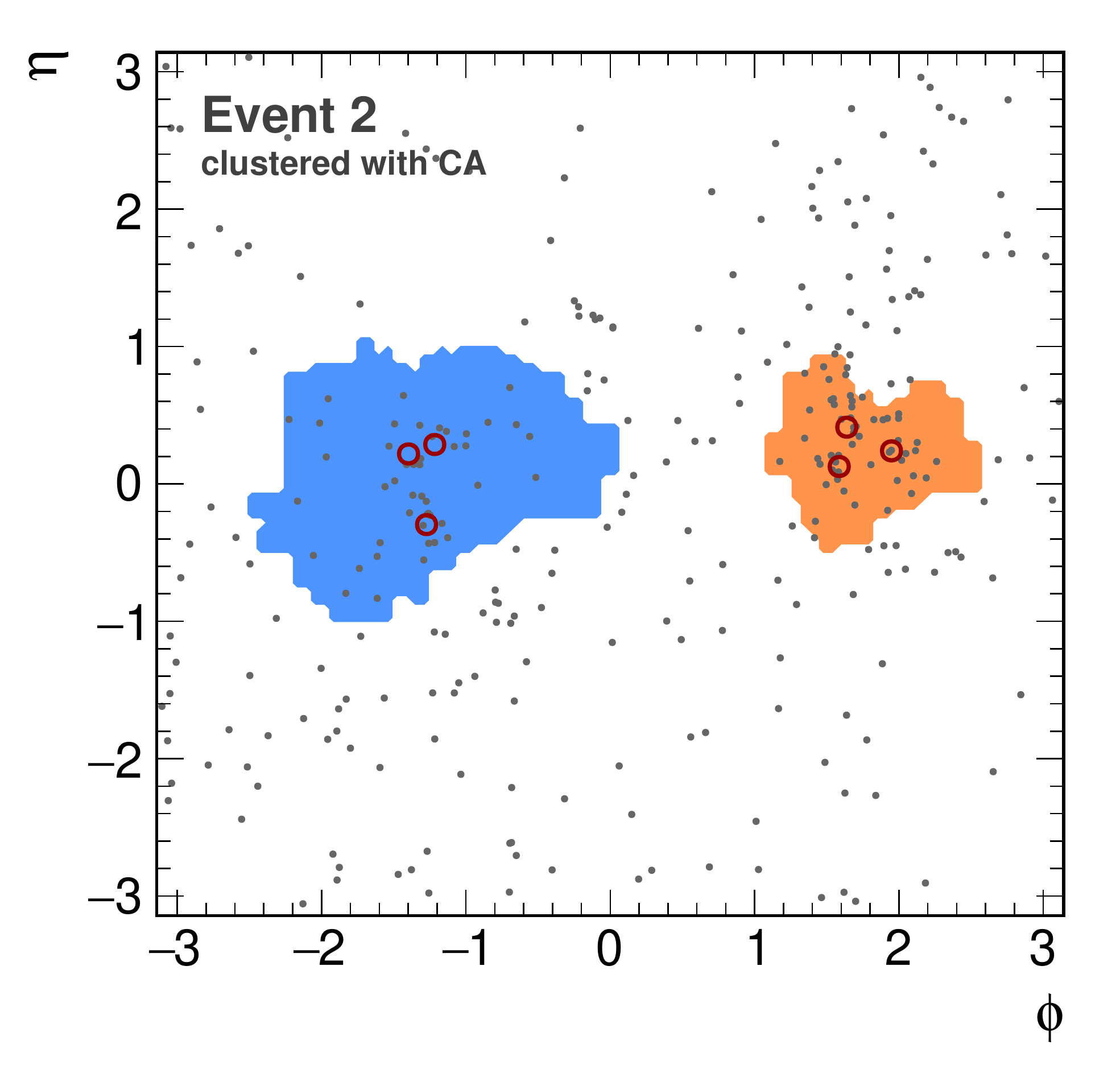}
  \includegraphics[width=0.47\textwidth]{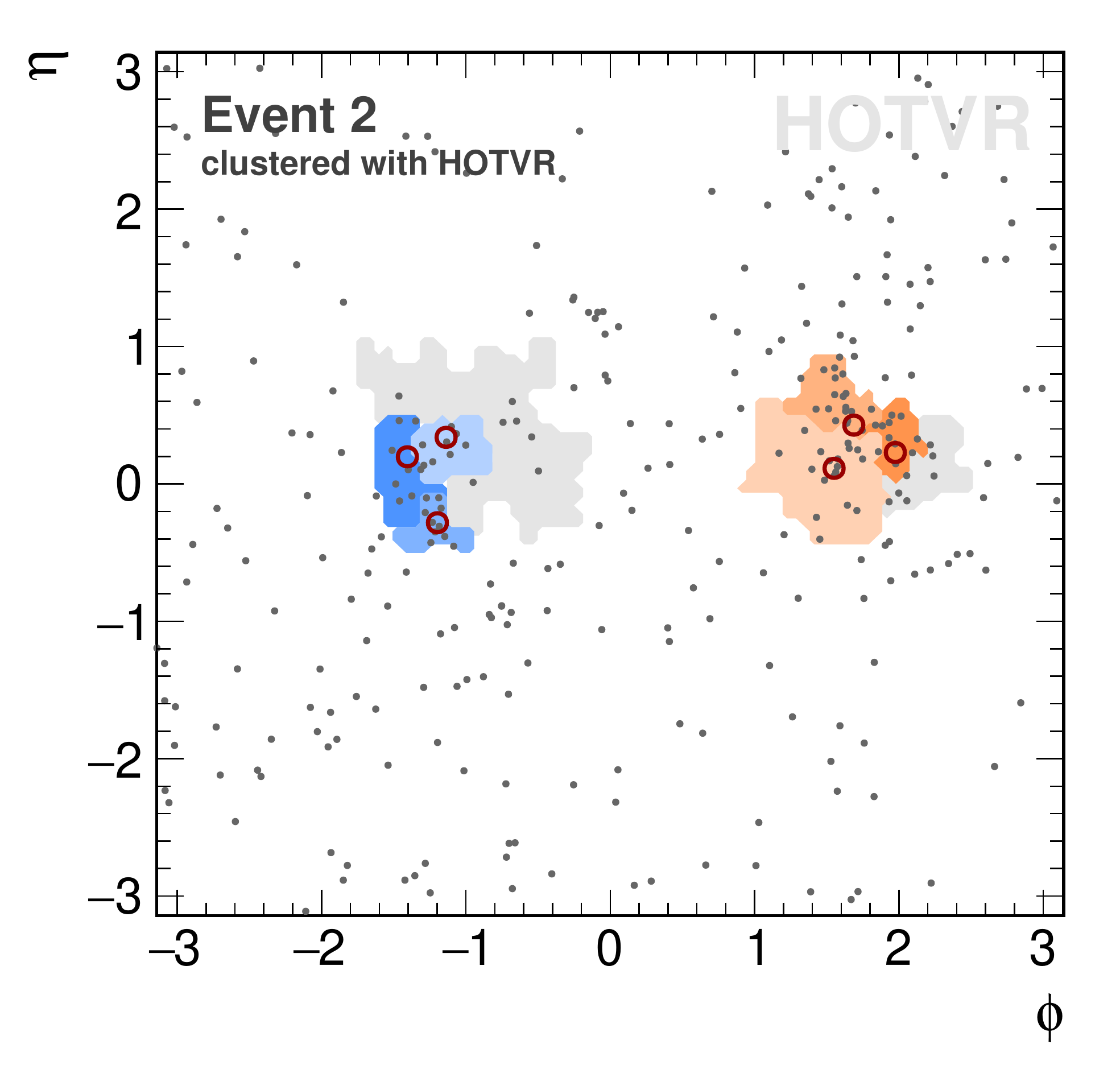}
  \caption{Two simulated \ttbar events clustered with the CA algorithm with distance 
  parameter $R=0.8$ (left column) and with the \HOTVR algorithm (right column). 
  The top quarks have either low \pt (top row, Event~1) or high \pt (bottom row, Event~2).
  The two leading jets in the events are shown as coloured areas (orange/blue).
  The stable particles, input for the jet finders, are drawn as grey dots. 
  The quarks from the top quark decay are depicted by red circles and are
  shown for illustration purposes only. In case of the \HOTVR algorithm
  the subjets are shaded from light to dark, corresponding to increasing \pt. The
  grey areas correspond to regions rejected by the mass jump criterion. 
  \label{fig:jetdisplay}}
\end{figure}
The algorithm's behaviour is visualised in Fig.~\ref{fig:jetdisplay}
where two example \ttbar events, generated with \textsc{Pythia\,8}~\cite{Sjostrand:2006za,Sjostrand:2007gs,Sjostrand:2014zea} at low \pt (top row, Event 1) and at high \pt (bottom row, Event 2), 
are clustered with the CA algorithm (left column) and with the \HOTVR algorithm (right column).
The active catchment areas of the hard jets are obtained using ghost
particles~\cite{Cacciari:2008gn} and are illustrated by the coloured (orange/blue)
areas\footnote{The exact borders of the jet areas depend
  slightly on the specific configuration of the ghost particles.}. 
The impact of the VR part of the algorithm is nicely illustrated by the largely different 
jet sizes of the two events clustered with the \HOTVR algorithm (right column).
The grey regions in the right panels were rejected
by the mass jump criterion and are not part of the \HOTVR jets. This criterion 
has largest impact in events at low \pt as exemplified in Event 1 (top, right).
The \HOTVR jets together with their subjets reproduce the kinematics of
the top decay adequately, both at low and high \pt, demonstrating a better adaptation to the
decay topology than CA jets. A similar picture is obtained when 
comparing \HOTVR jets to anti-\kt jets. 

\subsubsection*{Collinear and infrared safety} 

The \HOTVR algorithm is infrared and collinear (IRC) safe, except for
the unnatural parameter choice of $\mu=0$.  For parameter choices
corresponding to the original VR clustering, the \HOTVR algorithm is
trivially infrared and collinear (IRC) safe~\cite{Krohn:2009zg}. Similarly, for choices of
$\mu>0$ the algorithm is IRC safe, as soft and collinear splittings do
not generate mass.  This has also been verified in a numerical test, where the
stability of the jets as well as subjets found with the \HOTVR
algorithm was studied with respect to soft radiation and collinear
splittings. The algorithm proved to be IRC safe with no events out of
$10^{6}$ failing the test~\cite{tobiasphd}.

\subsubsection*{Timing} 

For timing tests, and throughout this work, the \Fastjet 3.2.1~\cite{Cacciari:2005hq, Fastjet} 
framework is used, together with \Fastjet Contribs version 1.024. 
Starting from \Fastjet version 3.2, advanced clustering strategies 
became available which led to substantial speed improvements, 
especially at high particle multiplicities. 
For this reason the run time of the algorithm has been 
studied for different particle multiplicity scenarios, 
low $\Order(50)$, medium $\Order(300)$ and high $\Order(3000)$.
In Tab.~\ref{tab:runtime} the CPU time of the \HOTVR algorithm with
default parameters (cf.\ Tab.~\ref{tab:VRhot_params}) is compared to
those of the CA jet algorithm~\cite{ca1,ca2}, 
the CMS top tagger~\cite{johnhopkins, cmstt}, 
the \HEPTT~\cite{htt2009, htt2010}, 
the \HEPTT in OptimalR mode~\cite{htt2015}, 
the VR algorithm~\cite{Krohn:2009zg} 
as well as the mass jump algorithm~\cite{Stoll:2014hsa}.
\begin{table}[tbp]
\centering
\small
\begin{tabular}{l c c c}
\multirow{2}{*}{Algorithm}               & \multicolumn{3}{c}{Particle Multiplicity}  \\
                                         &  $\Order(50)$ & $\Order(300)$  & $\Order(3000)$ \\ \hline
CA algorithm ($R = 0.8$)               &  $1.0$        & $1.0$          & $1.0$          \\
CA algorithm ($R = 1.5$)               &  $1.5$        & $1.3$          & $1.3$          \\
CMS top tagger ($R = 0.8$)             &  $1.3$        & $1.1$          & $1.1$          \\
\HEPTT ($R = 1.5$)                     &  $1.8$        & $1.3$          & $1.3$          \\
OptimalR \HEPTT ($R = 0.5\text{--}1.5$) &  $2.4$        & $1.4$          & $1.3$          \\
VR clustering ($\rho=600\gev$)           &  $1.3$        & $1.6$          & $5.8$         \\
Mass jump algorithm                      &  $1.7$        & $4.2$          & $23.8$        \\
\HOTVR                                   &  $1.6$        & $1.7$          & $5.3$          
\end{tabular}
\caption{CPU time comparison of various algorithms for low, medium and high particle 
multiplicities. The values are normalized to the CPU time of the CA algorithm with $R = 0.8$. 
\label{tab:runtime}}
\end{table} 
For the various top taggers the CPU time listed includes the time for the
underlying jet finding as well as for the top tagger specific processing steps.
The developments in \Fastjet 3.2 result in a much faster runtime of 
the VR and \HOTVR clustering, compared to previous versions (not shown). 
At low and medium multiplicities, the runtime of the \HOTVR algorithm is comparable 
to that of the other top-tagging algorithms tested. 
At high multiplicities, it is about a factor four slower than the \HEPTT 
algorithms, but it is still fast enough for practical uses\footnote{
  For example, on a MacBook Pro with a 2.5\,GHz Intel Core i5 processor and 16\,GB 1600\,MHz 
  DDR3 Memory the runtime is about 25\,ms per event for multiplicities of $\Order(3000)$. 
}.
The original mass jump algorithm has not been updated to employ the new clustering 
strategies, which leads to a much worse performance at medium and high multiplicities.

\section{Physics performance}
\label{sec:performance}

Studies of the physics performance are carried out using the 
event generator \textsc{Pythia~8}~\cite{Sjostrand:2006za,Sjostrand:2007gs,Sjostrand:2014zea}.
A $\pp\to\ttbar$ sample is used as signal process,  
background events are obtained by simulating QCD dijet production in \pp collisions.
For both samples a centre-of-mass energy of $\sqrt{s}=13$\tev is used,
the multiple parton interaction tune 
Monash 2013~\cite{Skands:2014pea} and
the LO NNPDF2.3~QCD+QED~\cite{nnpdf23} PDFs with $\alpha_s(M_Z) = 0.130$ 
are employed.
At this stage no additional \pp interactions during a single bunch 
crossing (pile-up) are simulated\footnote{ 
While pile-up effects will worsen the overall performance of 
the algorithm, the change is not expected to be significant for moderate
pile-up scenarios (up to 20--30 additional pile-up interactions).}. 

Throughout this work, jets are clustered using all stable particles from the 
\textsc{Pythia~8} output. 
In some studies, 
additional jets (labelled parton jets) are obtained using a list of 
all final state partons\footnote{
Final state partons are defined as partons which enter the hadronisation step.
} as input to the anti-\kt algorithm with distance parameter $R=0.4$ with a minimum \pt of 100\,GeV.
In case of \ttbar production, the top quark is effectively 
treated as stable for the purposes of defining the parton jet:  
after showering the top quarks are added to the parton list, and all partons from the 
top quark decay are removed.
In case a matching between particle and parton jets is employed, the 
geometrical matching condition $\Delta R < \Reff$ is used. 

\subsubsection*{Reconstruction of masses and transverse momenta}

\begin{figure}[tbp] 
  \centering
  \includegraphics[width=0.47\textwidth]{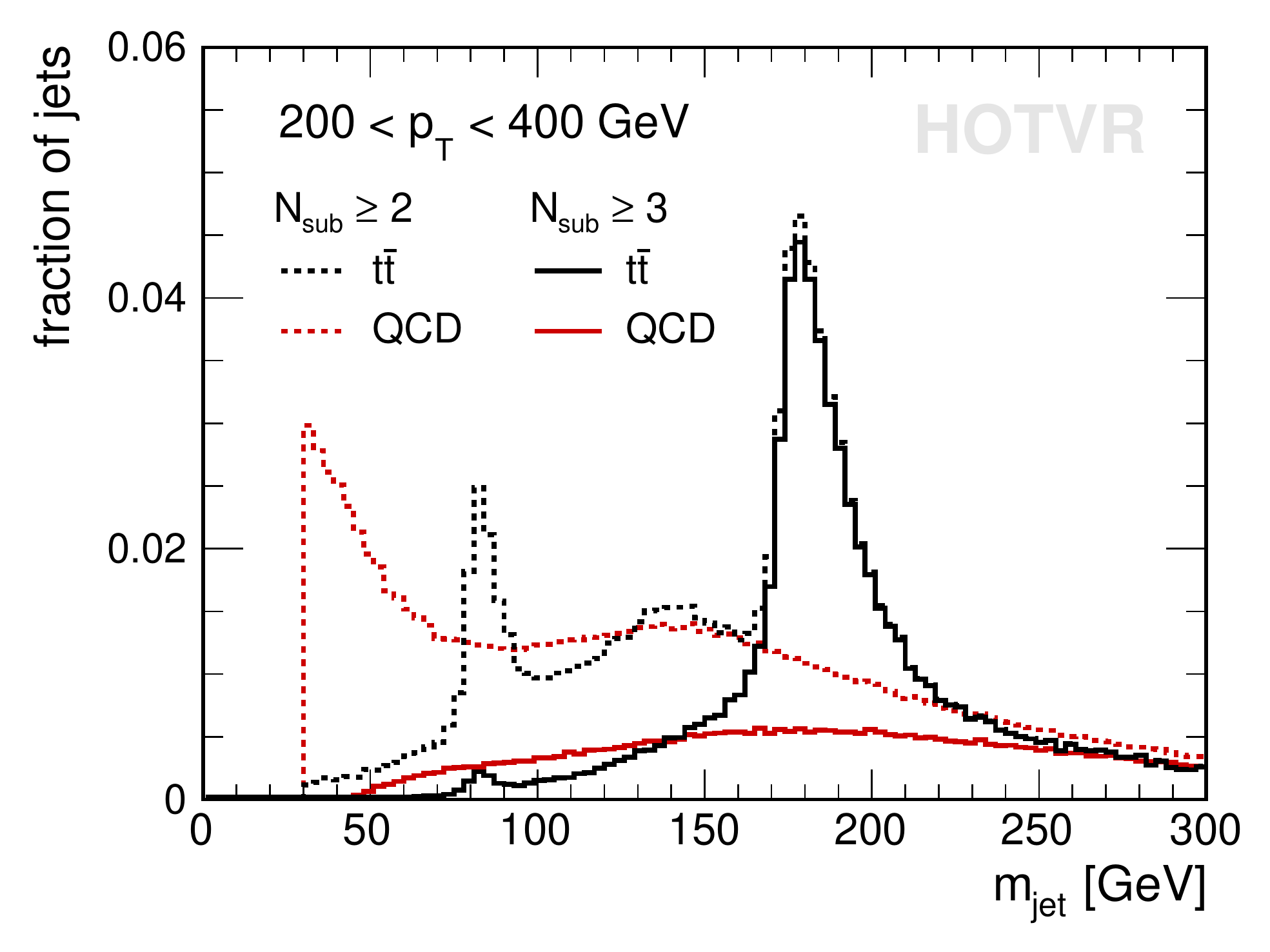}
  \includegraphics[width=0.47\textwidth]{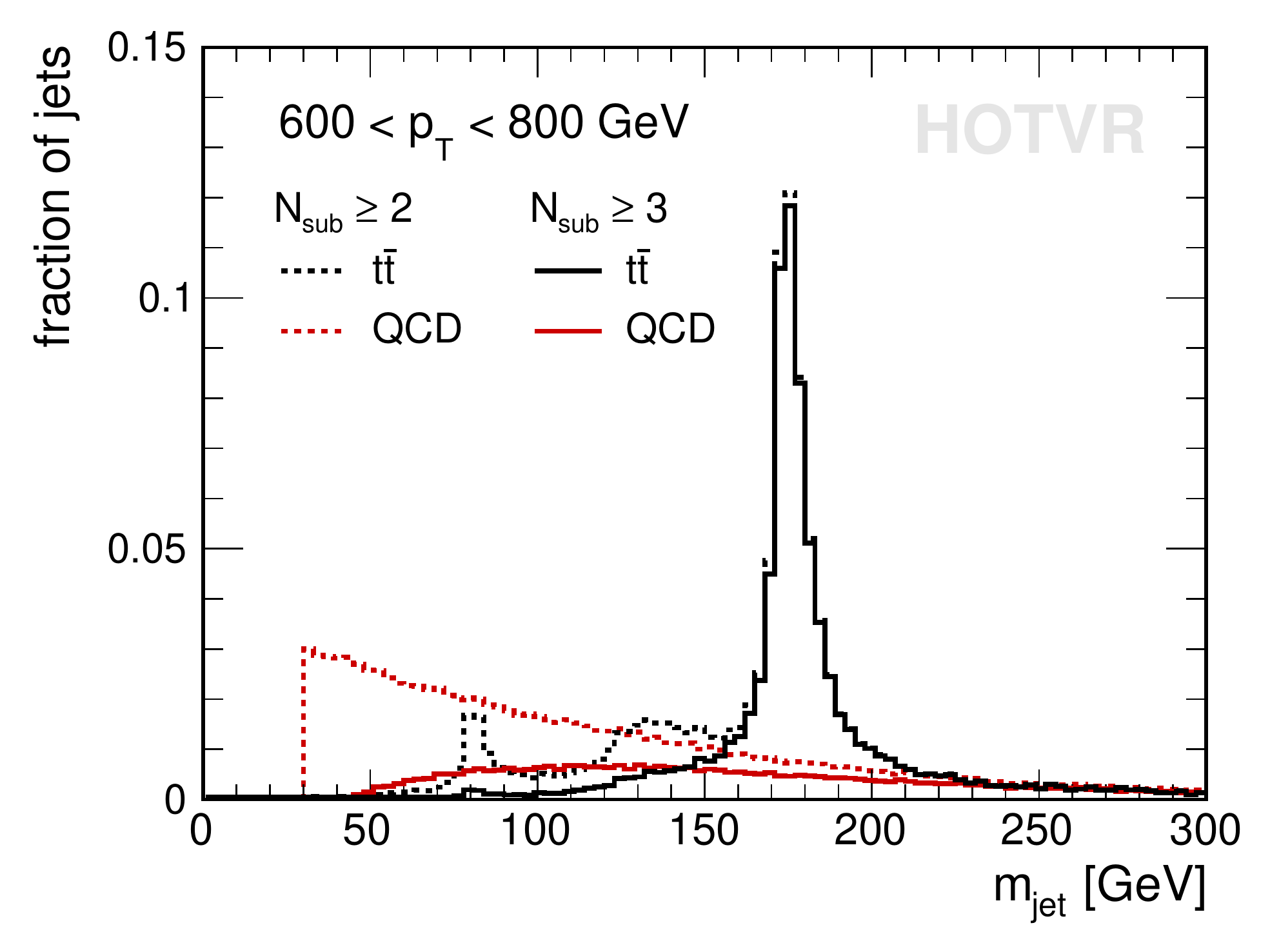} \\
  \includegraphics[width=0.47\textwidth]{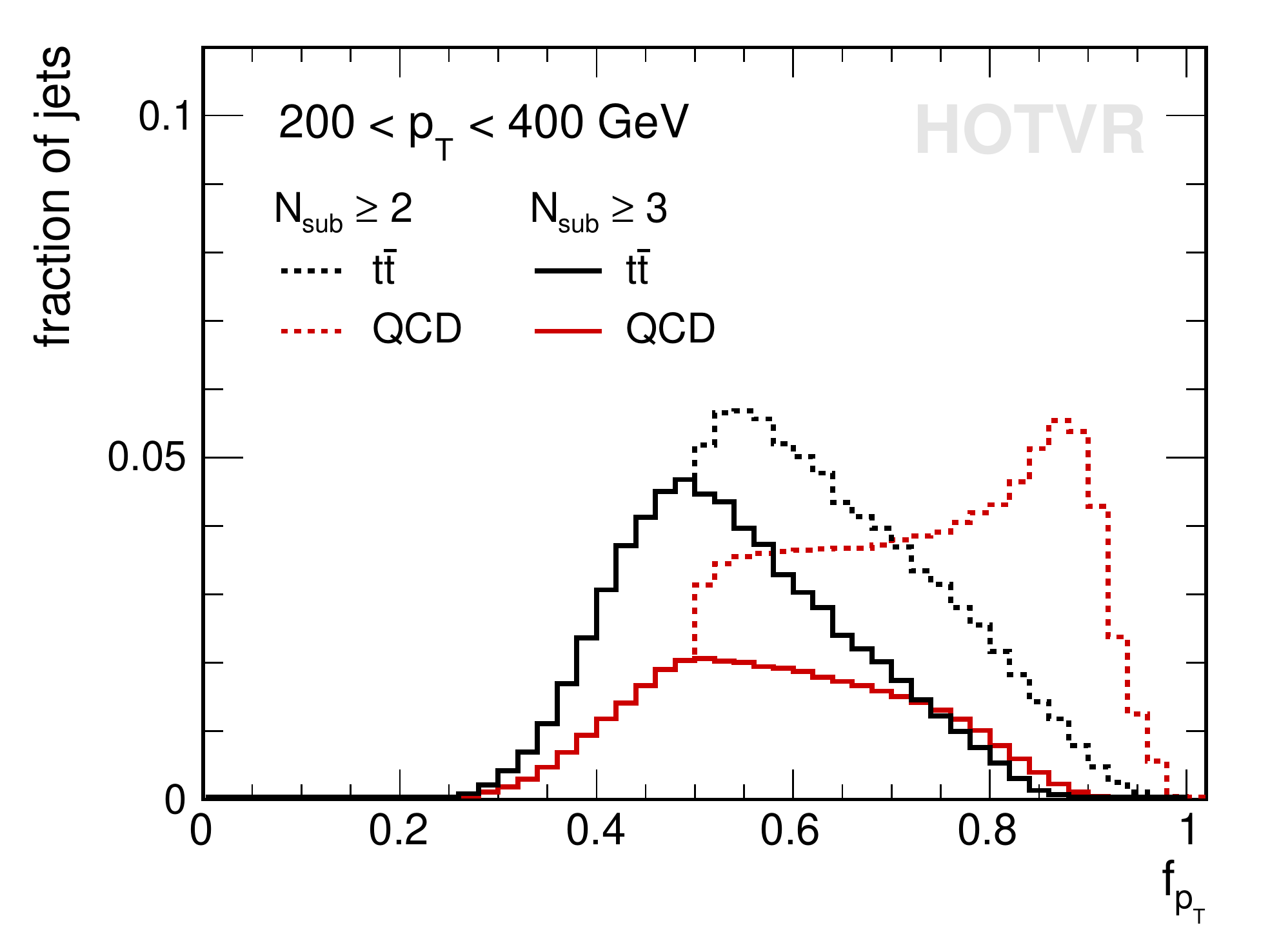}
  \includegraphics[width=0.47\textwidth]{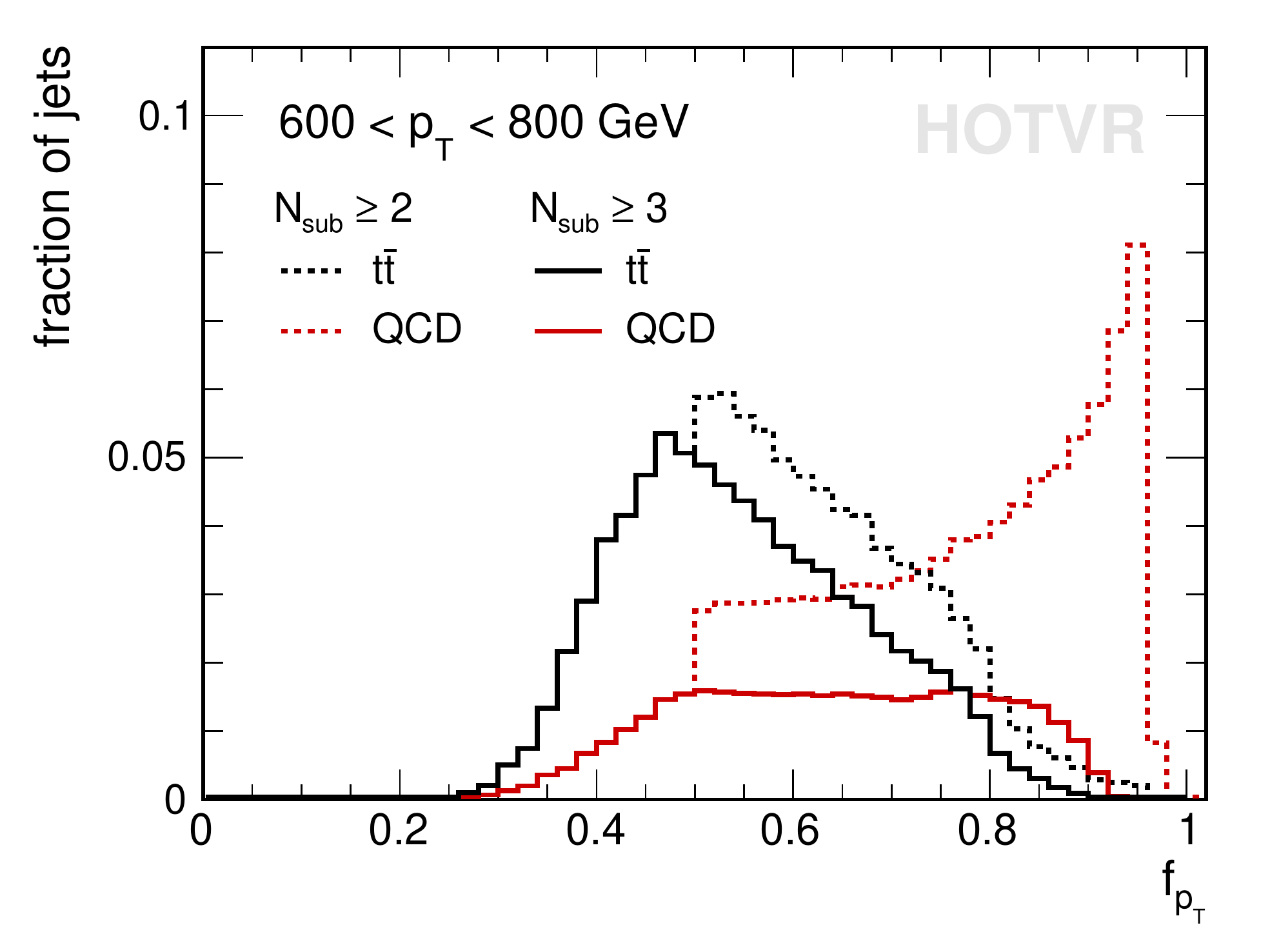} \\
  \includegraphics[width=0.47\textwidth]{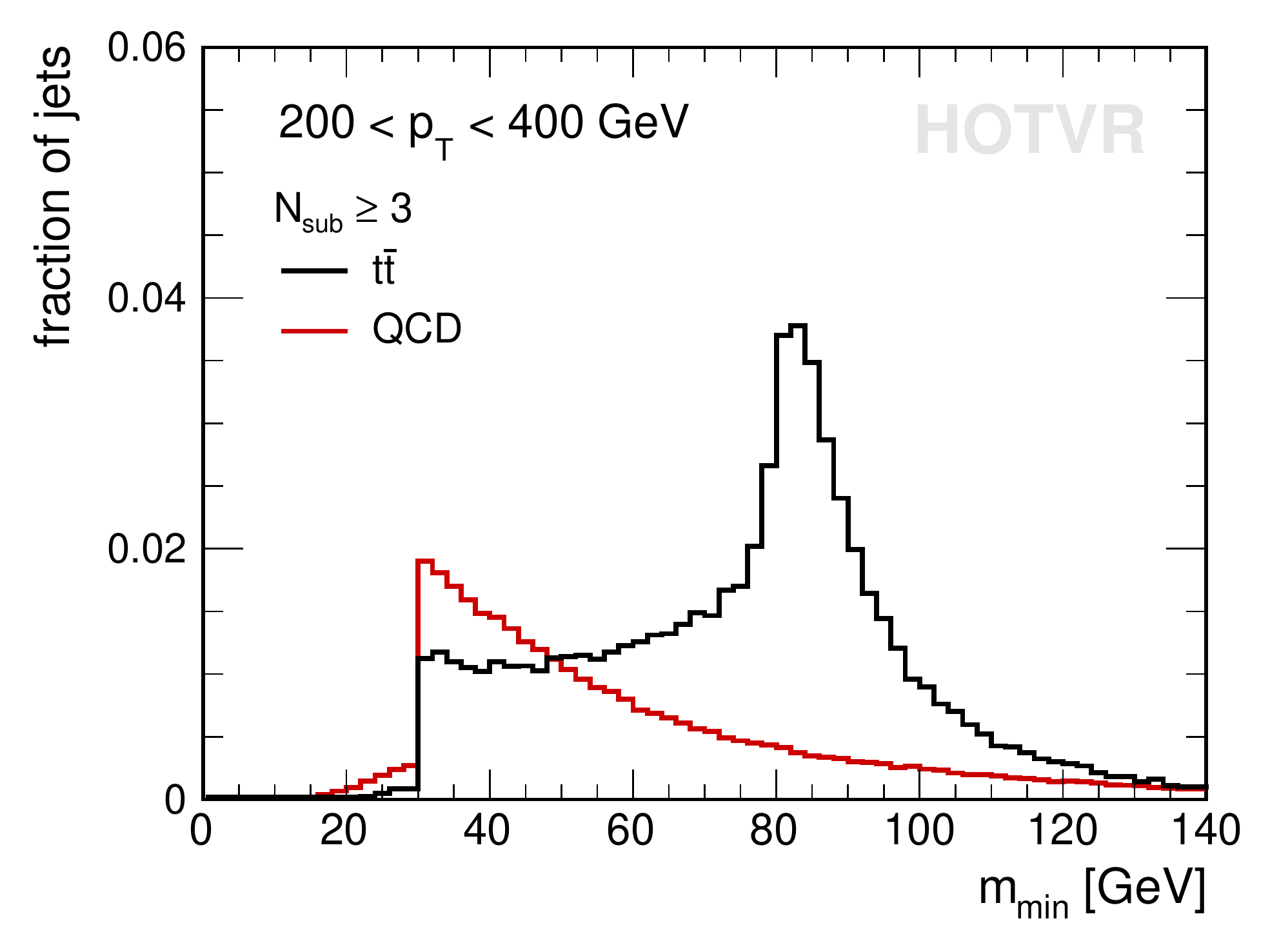}
  \includegraphics[width=0.47\textwidth]{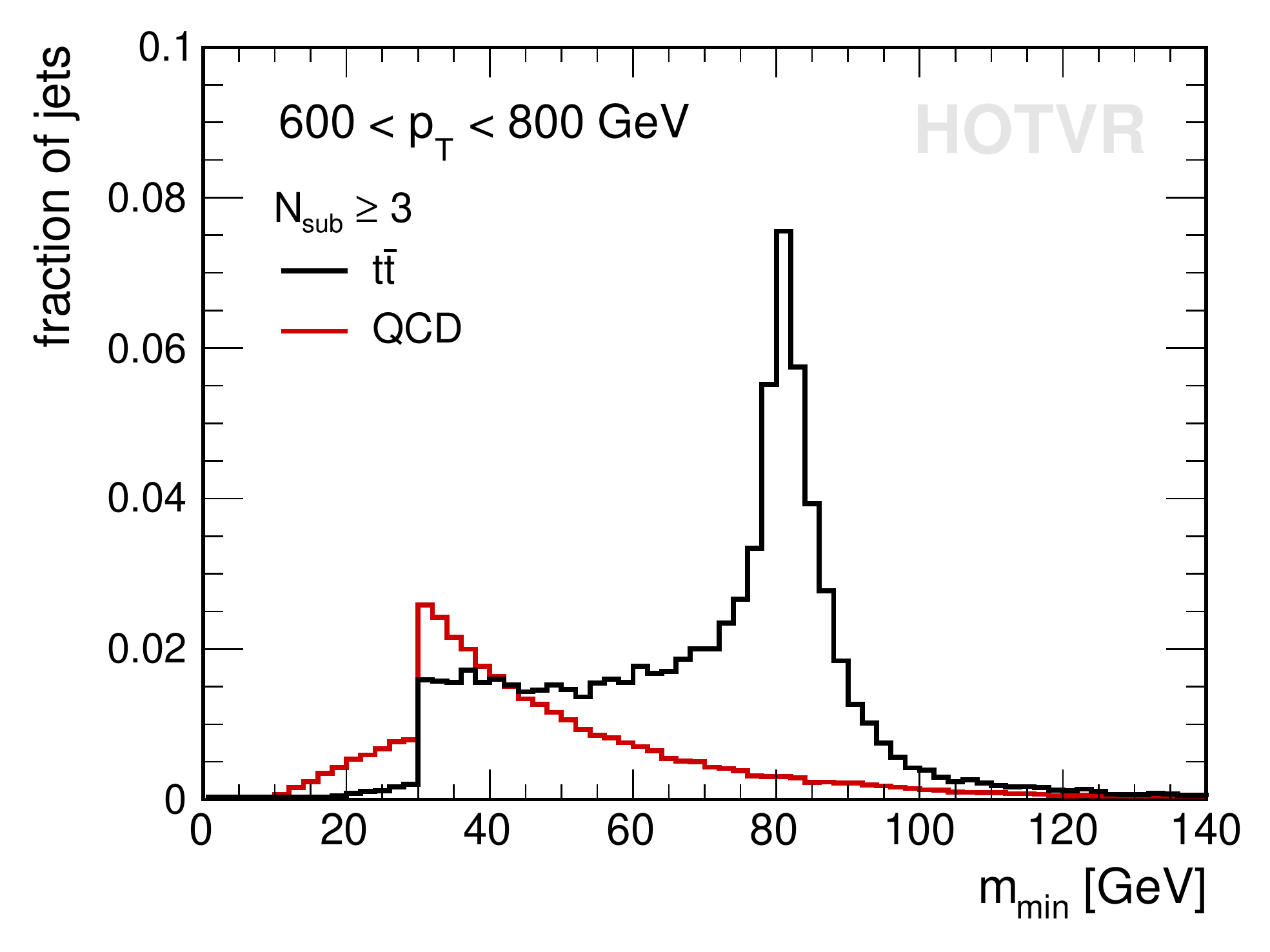}
  \caption{Distribution of the jet mass (top), 
  fractional leading subjet transverse momentum (middle)
  and minimum pairwise mass (bottom) 
  for signal (black) and background (red) events as obtained with the \HOTVR algorithm 
  for two different ranges in parton jet $\pt$. 
  The distributions are shown for subjet multiplicities $\Nsub \ge 2$ (dashed lines) 
  and $\Nsub \ge 3$ (solid lines). Note that the minimum pairwise mass is only 
  defined for $\Nsub \ge 3$. 
  The distributions have been normalised to unit area for $\Nsub \ge 2$. 
  \label{fig:hhtt_mass}}
\end{figure}

The key to the tagger's effectiveness is the accurate reconstruction
of subjets originating from the top quark decay, achieved by the VR 
condition and the mass jump criterion. This leads to a stable peak position 
for the mass of top jets over a large range of jet \pt, as 
shown in Fig.~\ref{fig:hhtt_mass}. The jet mass \mjet distribution 
for jets with two different subjet multiplicity \Nsub selections is 
shown for two ranges in the $\pt$ of the parton jet matched to the particle jet.
For $\ttbar$ events with $\Nsub \ge 2$ the distributions feature a dominant peak, 
stable around the top quark mass\footnote{
The VR algorithm alone affects the jet mass distribution 
similarly to a trimming~\cite{trimming} procedure
for anti-\kt jets at high top quark \pt~\cite{ATL-PHYS-PUB-2016-013}.
}
for fully merged decays, and two smaller peaks at lower masses 
corresponding to partially merged top quark decays. 
The requirement of $\Nsub \ge 3$ leads to a depletion of the two secondary peaks, 
while the peak around the mass of the top quark is hardly affected. 
At low \pt (left) the top quark peak is wider with a larger 
tail and is situated on a larger plateau than at 
high \pt (right) because of contributions from additional radiation which aggregate 
in the jet due to its large size. While this leads to a larger
misidentification rate at low \pt, it results in a non-vanishing efficiency
already at top quark transverse momenta as low as $100\gev$. 
For typical QCD jets a falling distribution is observed. 
The wide peak at mass values around 140\gev observed at low \pt (left) 
is a result of the subjet kinematics, where an angular separation of 
$\Delta R = 1.0\text{--}1.5$ leads to jet masses around this value. When changing 
the kinematics by relaxing the \ptsub requirement, the peak 
vanishes and a falling distribution is obtained. 
The width of this peak is reduced for intermediate ($400<\pt<600\gev$) 
transverse momenta (not shown) and a monotonically falling background distribution 
is obtained for values of $600<\pt<800\gev$ (right).
Very similar distributions are obtained for values of $\pt>800\gev$.

The distributions of the leading subjet's fractional transverse momentum 
$\fpt = \ptone/\pt$ is shown in Fig.~\ref{fig:hhtt_mass} (middle). Signal jets 
contain subjets with more evenly distributed transverse momenta, 
while for background jets the leading subjet carries a larger 
fractional \pt on average. 
The variable $\fpt$ shows good separation power between signal and background jets before 
a subjet multiplicity selection. After the requirement of $\Nsub \ge 3$,  
the separation power is reduced, but the variable is still useful, especially at high \pt.

For jets with $\Nsub \ge 3$, the distribution of the minimum
pairwise mass $\mmin$~\cite{johnhopkins, cmstt}, defined as the
minimum invariant mass of pairs of the three highest \pt subjets 
$\mmin = \min[m_{12},m_{13},m_{23}]$, is shown in Fig.~\ref{fig:hhtt_mass} (bottom)
for two regions of \pt of the parton jet.
The distributions show a clear cut-off at the chosen value of
the mass jump threshold ($\mu=30$\,GeV). Above this value the
distribution is steeply falling for background jets, while \ttbar
signal jets exhibit a pronounced peak around the value of the \W
boson mass, as expected for top quark jets. 
The tail below the mass jump threshold is a result of light subjets 
combined with a heavier pseudojet, fulfilling the mass jump criterion 
in step 3) of the algorithm.

Besides an adequate reconstruction of masses, algorithms should also
be able to reconstruct the kinematics of the initial heavy particle.
In particular the size of the catchment area, which is responsible for
the amount of additional radiation clustered into the jets, and the
intensity of the grooming procedure are critical components for the
performance in this area.  For an evaluation of the kinematic object
reconstruction by the \HOTVR algorithm, we calculate the \pt ratio of the \HOTVR
jets and the matched parton jets containing a top quark.  We find a mean
value of the \pt ratio of $1.0$ within small deviations of the order
of 1\%, independent of the parton jet \pt.  The widths of the \pt
ratio distributions are about 5\%.  This shows that the \HOTVR algorithm is
able to accurately reconstruct the kinematics of the heavy object 
with the parameter choice given above.

\subsubsection*{Selection cuts in top-tagging mode}

For the discrimination of hadronically decaying top quarks from QCD
multijets a selection based on simple cuts using commonly employed
substructure variables has been implemented.  The variables \mjet and \mmin
calculated from the \HOTVR subjets are in principle sufficient for
building a robust top tagger over a large region of \pt.  However, cuts on additional variables
have been added to obtain a selection that allows a fair
comparison with other top-tagging algorithms using similar
selections. 
To ensure only a limited impact of not-included experimental effects (e.g. broadening of distributions)
the cut values have not been optimised rigorously. Nevertheless, they 
result in an improved discrimination between
signal and background\footnote{A more sophisticated selection based
  on multivariate analysis techniques or more complex observables
  might provide further performance improvement~\cite{JME-15-002} over
  this simple approach. However, the aim of the studies presented here is a
  comparison of the performance of the \HOTVR algorithm with existing
  algorithms.}.
The following selection defines the standard working point of 
the \HOTVR algorithm in top-tagging mode.
\begin{enumerate}
\item The leading subjet is required to have a fractional transverse momentum with 
respect to the jet, $\fpt = \ptone/\pt<0.8$, which ensures that the 
jet's momentum is distributed among its subjets and not 
carried by only the leading subjet.
\item The number of subjets \Nsub
is required to be $\Nsub \ge 3$, which increases the probability of 
reconstructing fully merged top jets and rejects a fair amount of QCD jets.
\item The jet mass is required
to fulfil $140<\mjet<220\gev$.
\item The minimum pairwise mass has to fulfil
$\mmin > 50\gev$.
\end{enumerate}
These selection criteria lead to similar subjet kinematics as
obtained by the CMS and \HEPTT algorithms with default parameters. This provides the basis 
for the comparison made in the following.

\subsubsection*{Performance comparison with ROC curves}

The signal efficiency and misidentification rate are studied using
single variable receiver operating characteristic (ROC) curves.  
The signal efficiency \es is defined as the fraction of tagged jets matched
to parton jets containing the top quark, 
with respect to all top quarks decaying hadronically.  
The background efficiency (or misidentification rate) \eb is calculated as
the fraction of tagged jets matched to parton jets in a QCD multijet 
sample, with respect to the total number of parton jets.  
Both, \es and \eb 
therefore combine identification and matching efficiencies. 
These definitions allow for a comparison of different tagging algorithms, 
in particular using different choices of the jet
distance parameter $R$, since the reference $\pt$ is defined by the
parton jet matched to the tagged jet and does not depend on the
specifics of the tagging algorithm under study.

\begin{table}
\centering
\small
\begin{tabular*}{0.95\textwidth}{p{0.22\textwidth} | p{0.16\textwidth} p{0.22\textwidth} | p{0.26\textwidth} }
CMS top tagger         & \HEPTT                   &                  & OptimalR \\
$R = 0.8$            &  $R = 1.5$             &                  & $R = 0.5 \text{--} 1.5$ \\[0.1cm]\hline&\\[-0.45cm]
$\delta_p > 0.05$      &  $f_{\mathrm{drop}} = 0.8$ & $m_{23}/m_{123} > 0.35$               & same as \HEPTT         \\
$A = 0.0004$           &  $m_{\mathrm{cut}} = 30\gev$ & $0.2 < \arctan \frac{m_{13}}{m_{12}} < 1.3$             & $\Delta R = 0.1$  \\
$\Nsub \ge 3$          &  $\Rfiltmax = 0.3$           & $f_{W} = 0.15$             & $\mrecf - \mrec > 0.2 \mrecf$  \\
$\mmin > 50\gev$       &  $\Nfilt = 5$                & $140< m_{123} < 220\gev$             & $\Delta R_{\mathrm{opt}} < 0.5$  \\
$140< \mjet < 220\gev$ &  $\ptsub > 30\gev$           &                             &                                 
\end{tabular*}
\caption{\label{tab:tagger_wps} Settings of the top tagging algorithms used. The parameter $R$ 
is the distance parameter of the jet clustering. The definition of the parameters follows  
Ref.~\cite{JME-10-013} for the CMS top tagger, Ref.~\cite{htt2010} for the \HEPTT 
and Ref.~\cite{htt2015} for the \HEPTT in OptimalR mode.}
\end{table}
In the following the performance of the \HOTVR algorithm in
top-tagging mode is compared with the performance of three top-tagging
algorithms especially designed for dedicated regions of \pt: the CMS
top-tagger targets the region of high \pt, the \HEPTT is designed for
low \pt and its improved version with OptimalR has been developed to
extend its usability to higher \pt.  The free parameters of these
taggers are listed in Tab.~\ref{tab:tagger_wps} together with a choice
of working points~\cite{JME-10-013, htt2010, JME-15-002}.
The ROC curves are obtained by keeping the free
parameters fixed at the values given and scanning only the
N-subjettiness~\cite{Stewart:2010tn, Thaler:2010tr, Thaler:2011gf}
ratio $\ftau = \tau_{3}/\tau_{2}$ with 
$\beta=1$. The choice of $\ftau$ as scanning variable\footnote{The
  usual procedure for obtaining the ROC curves by scanning the free
  parameters of each algorithm could provide misleading results in
  this case, as it cannot be ensured that the usage of additional or different
  scanning variables for a given tagger would not improve its performance 
  considerably.
  } ensures an 
unprejudiced comparison of the algorithms, which all 
rely on different reconstruction techniques and substructure variables, since this variable 
is not used in the definition of any of the taggers under study.
Furthermore, $\ftau$ has been shown to improve the 
performance of existing taggers (see for example Refs.~\cite{JME-15-002,htt2015}). 

\begin{figure}[tb]
  \centering
  \includegraphics[width=0.47\textwidth]{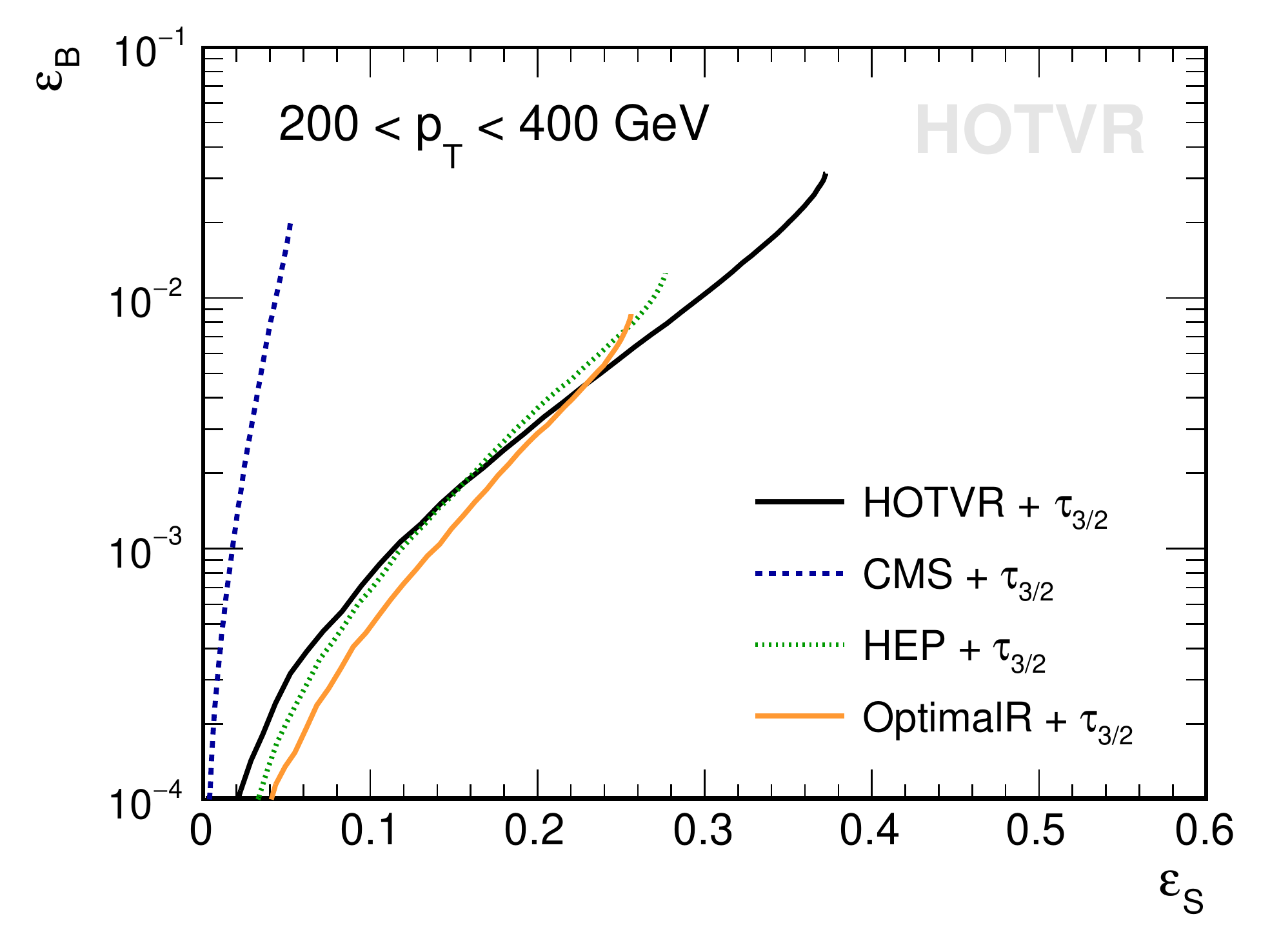}
  \includegraphics[width=0.47\textwidth]{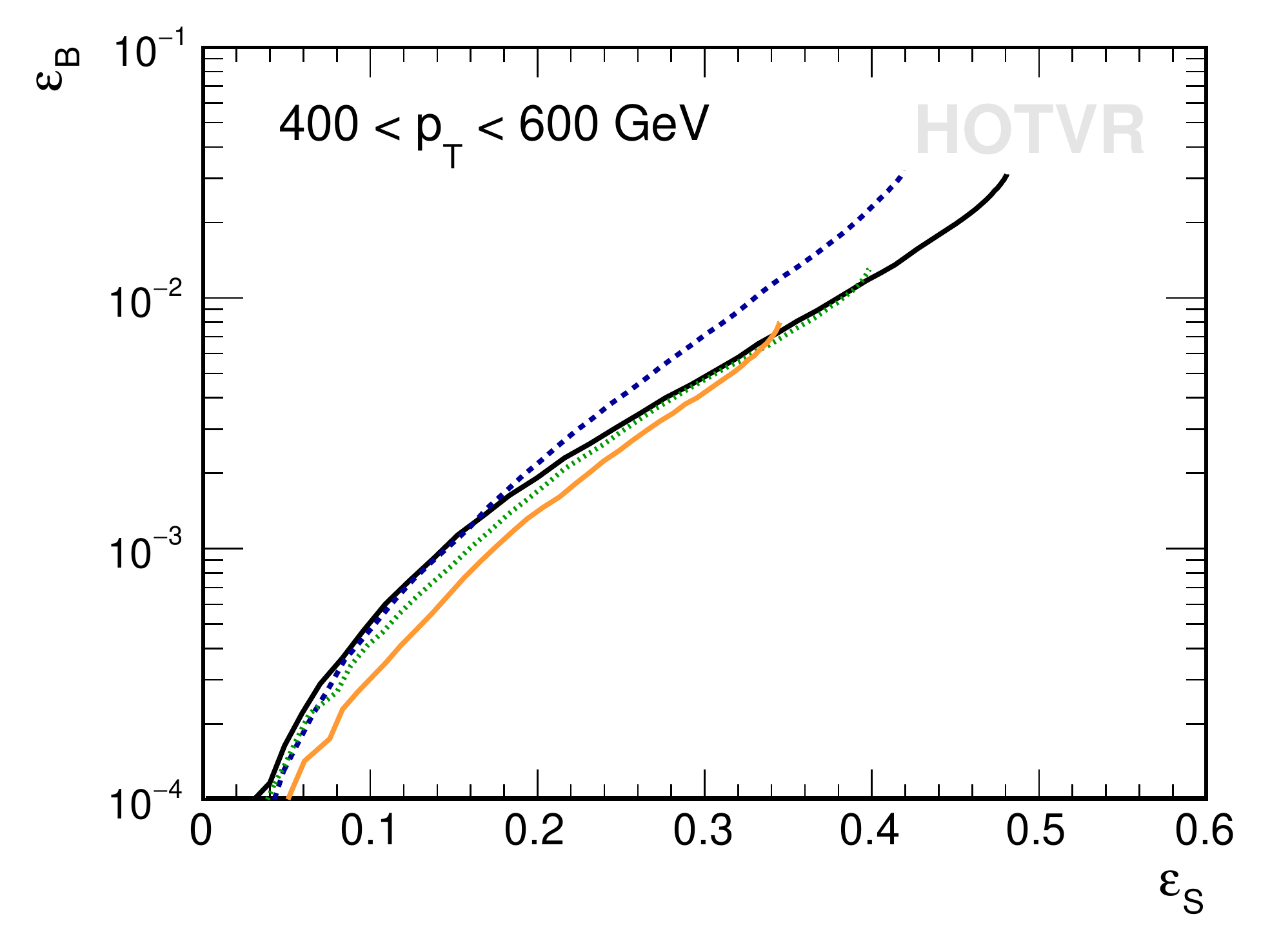}
  \includegraphics[width=0.47\textwidth]{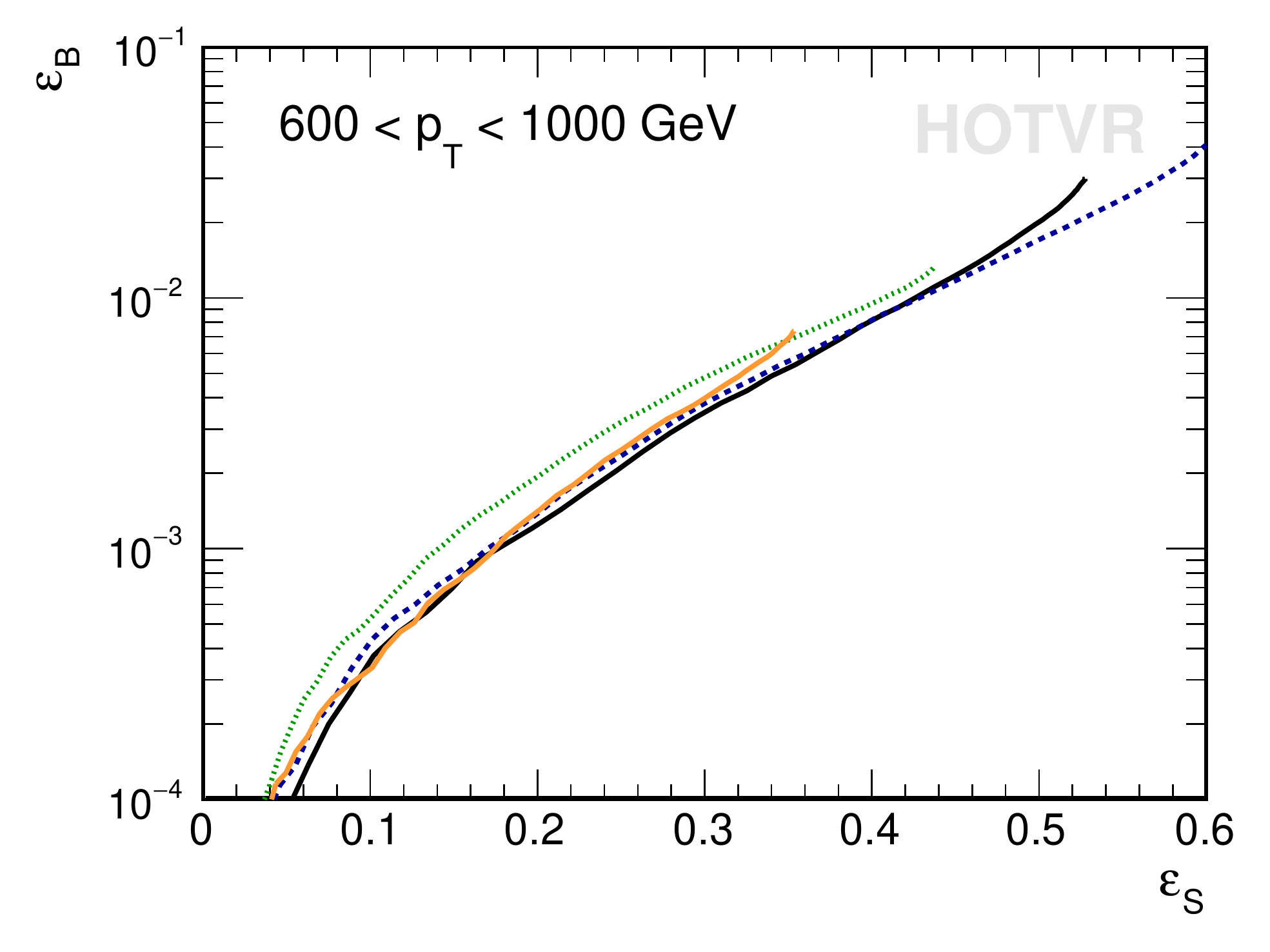}
  \includegraphics[width=0.47\textwidth]{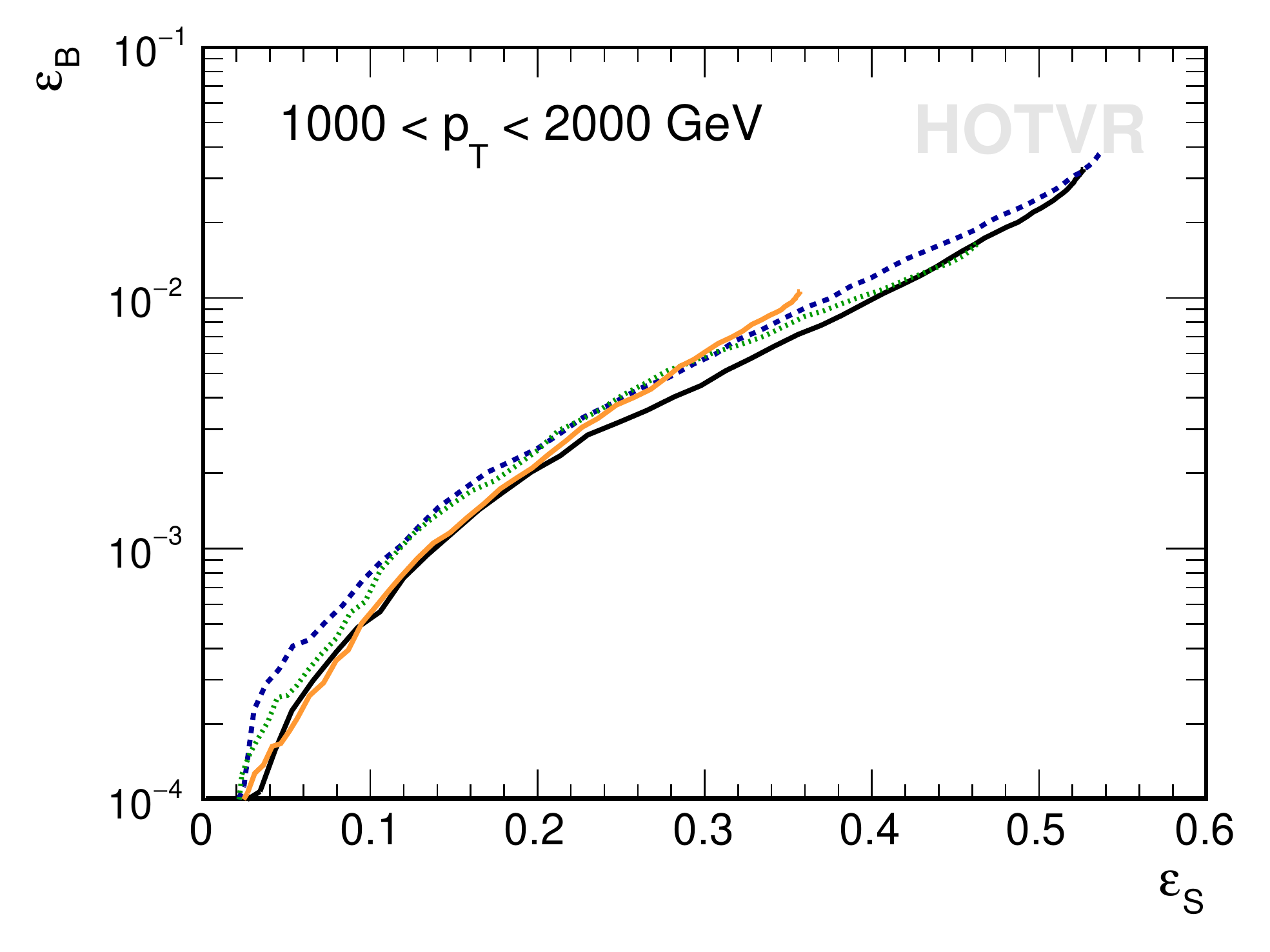}  
  \caption{Receiver operating characteristic (ROC) curves for different top tagging algorithms obtained from  
  a scan of the variable $\ftau$ in four different $\pt$ regions. 
  \label{fig:ROCs}}
\end{figure}
In Fig.~\ref{fig:ROCs} the ROC curves of the four top-tagging
algorithms are shown for four different \pt regions, where \pt is
defined by the parton jet matched to the tagged jet. The events
were reweighted to obtain a flat \pt spectrum such that all events in
the interval have the same weight.  At low \pt ($200<\pt<400\gev$, top
left) the CMS top-tagging algorithm has very small efficiency due to
the choice of $R = 0.8$ which results in jets not large enough to
cluster all particles from the top quark decay chain.  
The \HOTVR algorithm is able to provide a comparable performance as 
the two \HEPTT algorithms which were optimised for this \pt region.
For increasing values of \pt the
CMS tagger becomes more efficient, with a similar performance as 
the OptimalR \HEPTT starting from $\pt>600\gev$. 
In the \pt regions with $400<\pt<600\gev$ (top right) 
and $600<\pt<1000\gev$ (bottom left) 
the \HOTVR algorithm shows a similar relation between $\es$ and $\eb$ 
as the CMS and OptimalR \HEPTT, and is especially useful for 
high efficiencies.
In the highest \pt region considered ($1000<\pt<2000\gev$, bottom right)
the \HOTVR algorithm features overall the best performance over all $\es$ values, 
outperforming the CMS tagger, which was designed for the high \pt region. 

In summary, the \HOTVR algorithm shows a remarkably
stable performance over a large range in \pt with similar or even
better performance than algorithms especially designed for certain \pt
regions. Detector reconstruction and resolution effects, which are not included
in these studies, are expected to improve the performance 
of the \HOTVR algorithm relative to the other algorithms 
studied~\cite{tobiasphd}.

\section{Conclusion}
\label{sec:conclusion}

A new algorithm for the reconstruction and identification of
hadronically decaying heavy particles at the LHC has been introduced
in this paper.  The algorithm combines variable $R$ jet clustering 
with a veto based on a mass jump criterion. It performs jet
and subjet finding, and the rejection of soft radiation in one
sequence. This combination results in a stable determination of jet
substructure variables like the jet mass over a large range in \pt of
the heavy object. In top-tagging mode the \HOTVR algorithm provides an
excellent ratio of signal to background efficiency at low
top quark \pt as well as at high \pt, making the \HOTVR algorithm useful in the
regions of resolved and boosted decays at the same time.

While we focussed on top tagging in this work, the algorithm is also
applicable for the tagging of \W, \Z, \Hig or possible BSM resonances, where studies are ongoing.
Because of its algorithmic simplicity combined with remarkable performance,
this tagger could become a helpful ingredient for future boosted analyses
at the LHC.
 
\subsubsection*{Acknowledgements}
\label{sec:Acknowledgments}

\begin{details}
  We thank Michael Spannowsky for fruitful discussions during the 
  development of the algorithm. We also thank Jesse Thaler for 
  helpful suggestions on improvements of the document and for 
  advise on speed improvements. 
  This work is supported by
  the German Research Foundation (DFG) in the Collaborative Research
  Centre (SFB) 676 ``Particles, Strings and the Early Universe''
  located in Hamburg.
\end{details}

\addcontentsline{toc}{section}{References}
\bibliography{References}{}

\end{document}